\documentclass[apj]{emulateapj}
\usepackage{natbib}
\usepackage{amsmath}
\usepackage{apjfonts}

\newcommand{\dev}{\mathrm{d}}

\begin{document}

\title{Gamma-ray burst afterglow light curves from a Lorentz-boosted simulation frame and the shape of the jet break}

\author{Hendrik van Eerten$^1$, Andrew MacFadyen$^1$}
\affil{
  $^1$ Center for Cosmology and Particle Physics, Physics Department, New York University, New York, NY 10003\\
}

\begin{abstract}
The early stages of decelerating gamma-ray burst afterglow jets have been notoriously difficult to resolve numerically using two dimensional hydrodynamical simulations even at very high-resolution, due to the extreme thinness of the blast wave and high outflow Lorentz factors. However, these resolution issues can be avoided by performing the simulations in a boosted frame, which makes it possible to calculate afterglow light curves from numerically computed flows in sufficient detail to accurately quantify the shape of the jet break and the post-break steepening of the light curve. Here, we study afterglow jet breaks for jets with opening angles of 0.05, 0.1 and 0.2 radians decelerating in a surrounding medium of constant density, observed at various angles ranging from on-axis to the edge of the jet. A single set of scale-invariant functions describing the time evolution of afterglow synchrotron spectral break frequencies and peak flux, depending only on jet opening angle and observer angle, are all that is needed to reconstruct light curves for arbitrary explosion energy, circumburst density and synchrotron particle distribution power law slope $p$. These functions are presented in the paper. Their time evolutions change directly following the jet break, although an earlier reported temporary post-break steepening of the cooling break is found to have been resolution-induced. We compare synthetic light curves to fit functions using sharp power law breaks as well as smooth power law transitions. We confirm our earlier finding that the measured jet break time is very sensitive to the angle of the observer and can be postponed significantly. We find that the difference in temporal indices across the jet break is larger than theoretically anticipated and is about $-(0.5 + 0.5p)$ below the cooling break and about $-(0.25 + 0.5p)$ above the cooling break, both leading to post-break slopes of roughly about $0.25 - 1.3 p$, although different observer angles, jet opening angles and heuristic descriptions of the break introduce a wide range of temporal indices. Nevertheless, the post-break slope from our constant density ISM simulations is sufficiently steep to be hard to reconcile with post-break slopes measured for the \emph{Swift} sample, suggesting that \emph{Swift} GRBs mostly do not explode in a homogeneous medium or that the jet breaks are hidden from view by additional physics such as prolonged energy injection or viewing angle effects. A comparison between different smooth power law fit functions shows that although smooth power law transitions of the type introduced by Harrison et al. 1999 often provide better fits, smooth power law transitions of the type introduced by Beuermann et al. 1999 or even sharp power law fits are easier to interpret in terms of the underlying model. Light curves and spectral break and peak flux evolution functions will be made publicly available on-line at \url{http://cosmo.nyu.edu/afterglowlibrary}.
\end{abstract}


\section{Introduction}
\label{introduction_section}
Gamma-ray burst (GRB) Afterglows are produced by non-thermal radiation from collimated decelerating relativistic outflows following the collapse of a massive star or a neutron star-neutron star or neutron star-black hole merger (for reviews, see e.g. \citealt{Piran2004, Meszaros2006, Nakar2007, Granot2007}). Because they originate from cosmological distances, their jet nature can not be observed directly but is expected theoretically from constraints on the energy budget of the outflow (the isotropic equivalent energy of the afterglow often being comparable to the solar rest mass) and inferred observationally from the \emph{jet break} in the light curve. This break has been observed at various wavelengths ranging from radio to X-rays, and marks the onset of a steepening of the decay of the light curve. In this paper we present the most accurate description to date of the temporal and spectral evolution of the afterglow signal during the jet break, based on detailed relativistic hydrodynamics (RHD) calculations of the afterglow blast wave decelerating in a homogeneous medium.

The steeper decay following the break is the result of two changes in the outflow. On the one hand the jet is starting to become less collimated. As a result, the area of the blast wave front increases and the jet decelerates faster than before because it starts to sweep up more circumburst matter. On the other hand, the ongoing deceleration even without spreading reaches a point where the relativistic beaming cone of the synchrotron emission at and behind the shock front becomes sufficiently wide for the lack of flux beyond the edges of the jet to become visible, whereas before only a small patch along the direction of the observer could be observed and a jet was indistinguishable from spherical outflow. Both effects are expected to occur approximately around the same point in time, when jet half-opening angle $\theta_0 \sim 1 / \gamma$, with $\gamma$ the fluid Lorentz factor behind the shock. For the sideways spreading this is because $\theta_0 \sim 1 / \gamma$ marks the point where the fast spreading in the frame comoving with the jet becomes noticeable as well in the frame of the observer, while for the edge effect it marks the point where the beaming cones have become as wide as the jet itself. Although the widening of the jet was originally argued to be the stronger effect \citep{Rhoads1999}, subsequent numerical studies \citep{Granot2001, Kumar2003, Meliani2007, Zhang2009, vanEerten2011chromaticbreak, Wygoda2011, DeColle2012simulations} reveal the spreading not to be the exponential process described by \cite{Rhoads1999}, for observationally relevant jet opening angles. As a result, the edge effect plays a strong role in shaping the jet break and the angle of the observer relative to the jet axis becomes relevant even for observer angles within $\theta_0$ \citep{vanEerten2010offaxis, vanEerten2012observationalimplications, vanEerten2011hiddenswift}. By now, theoretical models incorporate more realistic descriptions of jet spreading \citep{Granot2012}.

Observational signatures of collimated afterglow outflow for a number of GRBs first started to emerge in the late nineties, both from the overall steepness of the light curve compared to theoretical expectations for a spherical explosion \citep{Sari1999} and observations of the jet break \citep{Beuermann1999, Fruchter1999, Harrison1999, Kulkarni1999}. Because of the complexity of the dynamics of decollimating jets, afterglow jet breaks have been modeled by heuristic functions for the purpose of data fitting. From the beginning connected power laws have been used by many groups (e.g. \citealt{Fruchter1999, Kulkarni1999}) but also power laws with a smooth transition (\citealt{Beuermann1999, Harrison1999}, using different descriptions for the transtion). Sharp and smooth power laws to describe jet breaks (or breaks in general) in afterglows have also been used in many more recent studies (e.g. \citealt{Zeh2006, Liang2009, Evans2009, Racusin2009, Guelbenzu2011, Oates2011, Fong2012, Panaitescu2012}). The advantage of a general heuristic function to describe afterglow breaks is that they do not necessarily assume an underlying model (i.e. \emph{jet} break) but aim to describe the observed shape of the data in a concise and convenient manner. Identifying breaks as jet breaks is a separate step, where the steepening of the break and pre- and post-break relations between spectral and temporal slope (the ``closure relations'', see e.g. \citealt{Zhang2004}) are compared against theoretical expectations. Using such methods a lack of afterglow breaks has been reported for the \emph{Swift} sample \citep{Kocevski2008, Racusin2008, Racusin2009}, which has been attributed to the quality of the data \citep{Curran2008} and the effect of the observer angle (\citealt{vanEerten2011hiddenswift}, providing a mechanism by which jet breaks can be postponed beyond \emph{Swift}'s capability to observe, as suggested by e.g. \citealt{Kocevski2008}). Recent light curves from numerical simulations demonstrate \citep{vanEerten2012scalings}  that the shape of the afterglow synchrotron spectrum changes strongly directly following the jet break, which renders the standard application of the closure relations unreliable and might serve to explain the lack of succes in using them to identify jet breaks \citep{Racusin2009}.

It has recently been pointed out \citep{vanEerten2012scalings} that the shape of the jet break in the light curve is determined by a scale-invariant function that depends only on initial jet opening angle $\theta_0$ and observer angle $\theta_{obs}$ and that scales in a straightforward manner between jet energies and between circumburst densities. This function is calculated from high-resolution relativistic hydrodynamics (RHD) simulations of the jet dynamics in 2D that include lateral spreading and deceleration to trans-relativistic velocities. 

This scale invariance has a number of useful practical implications. It makes it possible to distill a description of the jet break from numerical simulations that includes the full complexities of 2D trans-relativistic jet dynamics without the need to explicitly probe the parameter space in burst explosion energy and circumburst density with time-consuming RHD simulations. The resulting jet break description will be general and uniquely constrains the post-break closure relations and light curve slope. Existing smooth power law descriptions of the break can be compared against the simulation-derived shape. Simulation-derived jet break functions can even be fitted directly against the data in order to identify jet breaks. When these dimensionless jet break functions are scaled in order to fit the real time evolution of the data, the ratio $\rho_0 / E_{iso}$ is obtained, yielding important constraints on the physics of the progenitor.

Although significant progress has been made recently in numerically resolving afterglow jets properly using adaptive mesh refinement techniques \citep{Zhang2009, vanEerten2010offaxis, vanEerten2011chromaticbreak, Wygoda2011, vanEerten2012boxfit, DeColle2012simulations}, to date the early stages of the blast wave evolution have not been fully resolved due to the extreme sharpness of the blast wave profile in the self-similar Blandford-McKee (BM, \citealt{Blandford1976}) solution for an ultra-relativistic blast wave that provides the initial conditions for the simulations. As a consequence of this steepness, most matter in the blast wave is contained within a thin shell of typical width $\Delta R \sim R / 12 \gamma^2$, with $R$ the blast wave radius. When this thin shell is not completely resolved, this leads to a transient startup phase characterized by a temporary artificial drop in Lorentz factor of the outflow. Only after the blast wave has been evolved for some time does the fluid profile return to the shape predicted from analytically evolving the initial conditions. In practice, this transient feature would typically still be present at least at the onset of the jet break, because of the trade-off between decreased resolution at earlier starting times (due to the $\gamma$-dependency of the width) and decreased validity at late times (the closer to the jet break, the less valid the assumption of purely radial flow). In addition, when computing synthetic light curves, one has to account for the fact that the observed flux at a given point in observer time is made up of emission from a wide range of emission times, with contributions from the back of the jet being emitted earlier than those from the front in order to arrive at the same time. 

In order to obtain a truly accurate picture of the shape of the jet break, it is therefore required to resolve the pre-break flow of the blast wave completely up to sufficiently high Lorentz factor that the effect of the start-up transient is removed. In the current study we completely resolve the afterglow blast wave at extremely high Lorentz factors and early times by performing the RHD calculation in a different frame than the usual burster frame, which is at rest with respect to the explosion engine and the observer (aside from a cosmological redshift). By changing to a frame moving at fixed relativistic velocity along the jet axis, the narrowness of the jet profile due to Lorentz contraction is reduced and all relative Lorentz factors become small \citep{MacFadyen2013}. The price that is paid for this frame transformation, the loss of simultaneity across the grid, can be accounted for when the radiation from the evolving blast wave is calculated.

The features of the dynamics of narrow jets and ultra-high initial Lorentz factor ($\ge 100$) flows will be presented in a separate study \citep{MacFadyen2013}. In this work we limit ourselves to the radiation from afterglow jets and determine the general shape of the jet break for afterglow blast waves that start out highly relativistic (Lorentz factor of 100) and have an initial half-opening angle $\theta_0$ of 0.05, 0.1 or 0.2 rad., moving into a homogeneous environment. The observer angle is varied from observers looking straight into the jet to observers positioned on the edge of the jet. In \S \ref{dynamics_section} we discuss the methods of our RHD simulations and our implementation of the BM initial conditions in a boosted frame. In \S \ref{radiation_section} we discuss how light curves are calculated from simulations. In \S \ref{criticals_section} we show our results for the small set of key characteristic quantities (i.e. the break frequencies of the power law synchrotron spectrum and the peak flux) that determine the afterglow spectrum. We then use the characteristic quantities to calculate afterglow light curves at optical and X-ray frequencies in \S \ref{lightcurves_section} and compare the shape of the jet break to earlier parametrizations from the literature. Our results are summarized and discussed in \S \ref{summary_section}. Some technical aspects concerning radiative transfer from a Lorentz-boosted simulation frame are discussed in appendix \ref{boosted_frame_appendix}.

\section{Methods for blast wave dynamics}
\label{dynamics_section}

We assume that the radiation and the dynamics of the collimated relativistic blast wave can be separated. This assumption remains valid as long as the emitted energy is only a small fraction of the blast wave energy and as long as there is neglible feedback from the radiation on the jet dynamics. Additionally, we assume that the magnetic fields generated  at the front of the blast wave also contain only a small fraction of the available energy. Under these assumptions the jet dynamics can be computed using RHD simulations.

\subsection{Description of the RHD code}
\label{numerics_subsection}

We employ the \textsc{ram} parallel adaptive-mesh refinement (AMR) code \citep{Zhang2006}. The AMR technique, where the resolution of the grid can be dynamically doubled locally where necessary, is important in order to resolve the wide range of spatial scales involved, given the $\Delta R \sim R / 12 \Gamma^2$ width of the blast wave in the lab frame in which the origin of the explosion is at rest, where $\gamma$ can be $> 100$ for a typical afterglow blast wave. \textsc{Ram} makes use of the PARAMESH AMR tools \citep{MacNeice2000} from FLASH 2.3 \citep{Fryxell2000}. We use the second order F-PLM scheme \citep{Zhang2006} for the hydrodynamical evolution. In this study, a Taub equation of state is used where the adiabatic index smoothly varies between 4/3 for relativistic fluids and 5/3 for non-relativistic fluids, as a function of the ratio between comoving density and pressure \citep{Mignone2005, Zhang2009}:
\begin{equation}
 (h - 4p)(h - p) = \rho^2,
\end{equation}
where $p$ the pressure, $\rho$ the comoving density and enthalpy $h = \rho c^2 + p + e$, with $e$ the energy density.

\subsection{Scale invariant initial conditions}
\label{initial_conditions_subsection}

Blast waves for three different jet half opening angles were simulated for this study: $\theta_0 = 0.05$, $0.1$ and $0.2$ rad. The circumburst number density $n_0$ of the interstellar medium (ISM) is kept fixed at $1$ cm$^{-3}$, or $\rho_0 = m_p $ g cm$^{-3}$, with density $\rho$ and number density $n$ related according to $\rho \equiv n \times m_p$, where $m_p$ the proton mass. A more general expression for the circumburst density environment is given by$\rho_0 \equiv \rho_{0,ref} (r / r_{ref})^{-k} \equiv A r^{-k}$, with $r$ the radial coordinate, $\rho_{0, ref}$, $r_{ref}$ and $A$ parameters setting the density scale and $k$ setting the power law slope of the medium. Boosted frame simulations of blast waves decelerating in a stellar wind environment where $k = 2$  will be presented in a follow-up study. All simulations start from the self-similar BM solution for impulsive energy injection with isotropic equivalent explosion energy $E_{iso}$ set at $10^{53}$ erg. The actual values for the initial energy and circumburst density are completely arbitrary and the hydrodynamics equations can be expressed in terms of dimensionless variables. Generalizing these variables from the ISM case \citep{vanEerten2012boxfit} to arbitrary $k$ values, we have
\begin{equation}
 \mathcal{A} = \frac{r}{ct}, \quad \mathcal{B} = \frac{E_{iso} t^2}{A r^{5-k}}, \quad \theta, \quad \theta_0,
\end{equation}
that are scale invariant under the transformations
\begin{eqnarray}
 E'_{iso} & = & \kappa E_{iso}, \nonumber \\
 A' & = & \lambda A, \nonumber \\
 r' & = & (\kappa / \lambda)^{1/(3-k)} r, \nonumber \\
 t' & = & (\kappa / \lambda)^{1/(3-k)} t.
\end{eqnarray}
All scale-invariance relations follow from straightforward dimensional analysis, and are therefore not limited to the ultra-relativistic self-similar BM solution but apply throughout the evolution of the blast wave when jet spreading and deceleration occur.

\subsection{Simulations in a boosted frame}

Two challenging aspects of numerically simulating BM type outflows are the severe steepness of the radial profile of the various fluid quantities (i.e. the primitive quantities Lorentz factor $\gamma$, comoving density $\rho$, pressure $p$, and consequently the conserved quantities as well) and the ultra-relativistic nature of the outflow. Resolution issues regarding numerically calculated blast waves and light curves have been discussed by various authors \citep{Zhang2009, vanEerten2010transrelativistic, vanEerten2012boxfit, DeColle2012simulations}. As mentioned in the introduction, the most striking feature of an underresolved BM blast wave is a temporary spurious drop in Lorentz factor near the shock front. Because the observed flux strongly depends on the Lorentz factor ($F_\nu \propto \gamma^2$, due to relativistic beaming), this strongly impacts the light curve. In order to understand the early time dynamics, it is important to start from a time when outflow peak Lorentz factor $\gamma \gg 1 / \theta_0$ (the point at which sideways spreading is expected to become relevant and when the edges of the jet become observable) and ideally, any transient behavior due to numerical resolution should have subsided before this point.

In the current work we have used cylindrical ($R$, $z$) coordinates. The initial conditions were provided by the BM solution \citep{Blandford1976}, but expressed in a Lorentz boosted frame \citep{MacFadyen2013}. For all simulations in this study, the simulation frame was boosted with Lorentz factor $\gamma_S = 5$. All jets start with peak lab frame Lorentz factor $\gamma_0$ of 100 at the on-axis tip of the jet, though some were also run with $\gamma_0 = 50$ to check for convergence (which is expected for $\gamma_0 \gg 1 / \theta_0$).

\subsection{Resolution}

\begin{figure}
 \centering
  \includegraphics[width=\columnwidth]{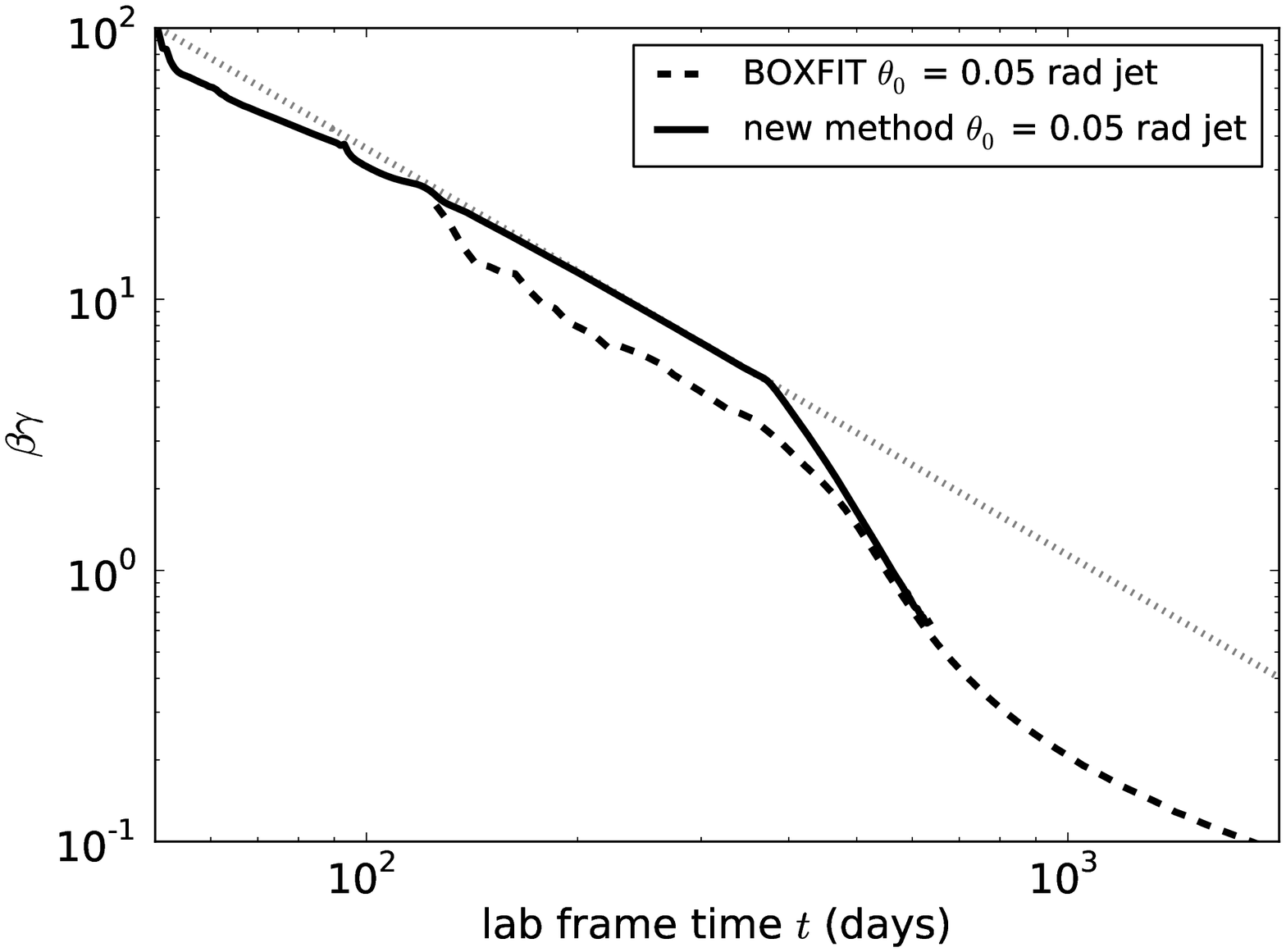}
  \includegraphics[width=\columnwidth]{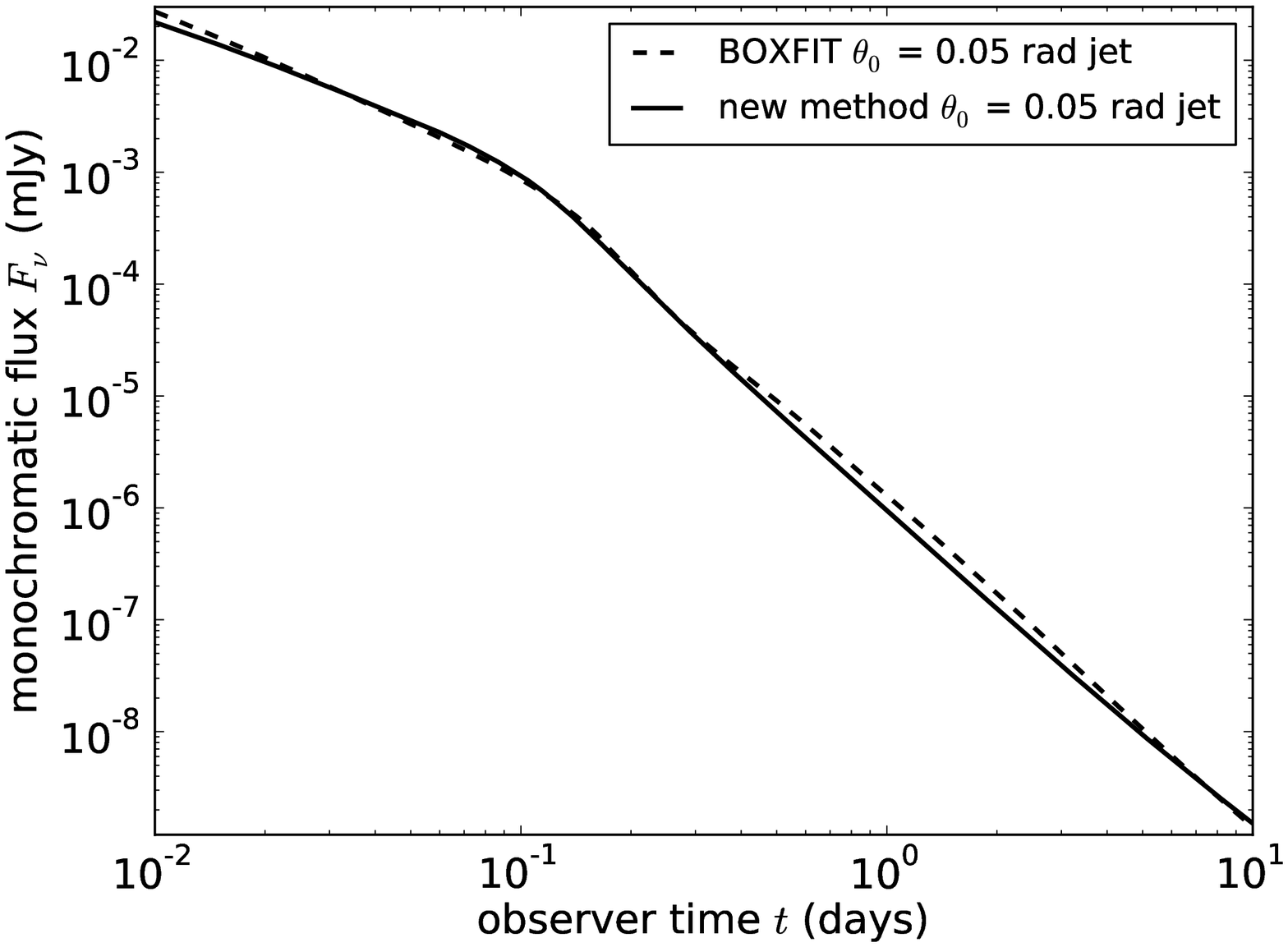}
\caption{Comparison of jet ($\theta_0 = 0.05$ rad) resolution between \textsc{boxfit} (dashed curve) and moving frame (solid curve) for a jet moving into the ISM. Top figure shows the evolution of the peak blast wave Lorentz factor (using $\beta \times \gamma$, the spatial component of the four-velocity) along the jet axis. According to the BM solution, $\gamma \propto t^{-3/2}$ and this slope is indicated by a dotted line. The bottom plot shows an X-ray light curve with jet break observed at $5 \times 10^{17}$ Hz, calculated using $E_{iso} = 10^{53}$ erg, $n_0 = 1$ cm$^{-3}$ (ISM), $p = 2.5$, $\epsilon_e = 0.1$, $\epsilon_B = 10^{-3}$, $\xi_N = 1$.}
 \label{resolution_figure}
\end{figure}

The simulation frame time duration of each simulation was $2 \times 10^7$ seconds. The cylindrical grids run from $-2 \times 10^7$ ls (lightseconds) to $2 \times 10^7$ ls in the $z$ direction and out to $2 \times 10^7$ ls in the $R$ direction perpendicular to the jet axis. The initial peak refinement level is 15, with 8 base level blocks and 8 cells per block in each direction. The smallest cell size at peak refinement level is therefore $\delta z = 2 \delta R = 19.1$ ls $ = 5.72 \times 10^{11}$ cm. Note that these are expressed in the boosted frame, so that the resolution in the $z$-direction is better compared to the lab frame\footnote{Throughout this paper we will use ``lab frame'' to refer to the frame in which the origin of the explosion and the unperturbed interstellar medium are at rest.} by a factor of $\gamma_S$. We enforce an upper limit on the total number of blocks on the grid. As the blast wave expands in size on the grid, the peak refinement level is decreased in order not to exceed this block limit.

The top panel of Fig. \ref{resolution_figure} shows the evolution of the Lorentz factor at the shock front (in the lab frame) along the jet axis, compared to that in earlier work \citep{vanEerten2012boxfit}, for the case where $\theta_0 = 0.05$ rad. The Lorentz factor is measured at the numerically determined momentum maximum, which is slightly behind the exact position of the shock front.  The Lorentz factor of the boosted frame simulation (solid line) agrees with the BM solution at that position to within $\sim 1\%$.  The dashed line shows the BM scaling $\gamma \propto t^{-3/2}$ appropriate for the behavior of the shock Lorentz factor. 
 
\section{radiation}
\label{radiation_section}

\begin{figure}
 \centering
  \includegraphics[width=\columnwidth]{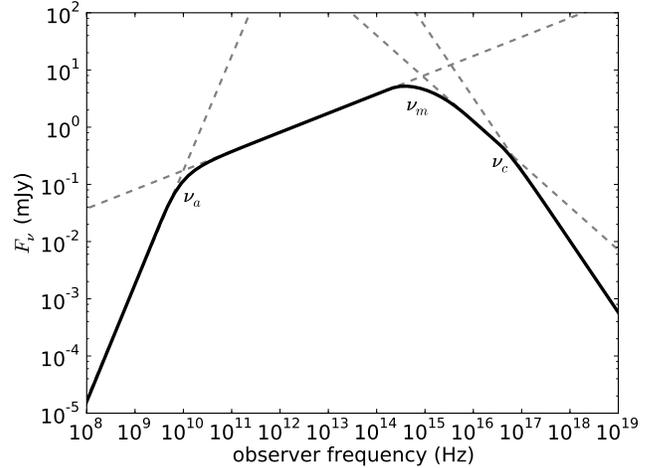}
\caption{Early time (observer time $t = 10^{-2}$ days) pre-break spectrum. $E_{iso} = 10^{53}$ erg, $n_0 = 1$ cm$^{-3}$ (ISM), $p = 2.5$, $\epsilon_e = 0.1$, $\epsilon_B = 10^{-3}$, $\xi_N = 1$. Dashed grey lines indicate asymptotic slopes, from left to right: $2$, $1/3$, $(1-p)/2$, $-p/2$.}
 \label{powerlaws_figure}
\end{figure}

The algorithm used to calculate the radiation for a given observer time, frequency, angle and distance is nearly identical to that used in \cite{vanEerten2012boxfit}, which in turn was based on \cite{Sari1998} and \cite{Granot1999}. The only difference is that it is now applied to a boosted frame simulation rather than a non-moving frame simulation. The radiative transfer equations are solved for a large number of rays through the evolving blast wave. The stepsize along the rays is set by the number of data dumps from the simulation (3000, although we found in practice that the light curves were converged even for 300 data dumps). The conceptual details of the radiative transfer approach for a moving frame are provided in appendix \ref{boosted_frame_appendix}.

The emission and absorption coefficients are calculated for synchrotron emission and synchrotron self-absorption (s.s.a.). The local synchrotron emission spectrum is given by a series of sharply connected power laws, with peak flux and spectral breaks determined by the local state of the fluid.  A relativistic distribution of shock-accelerated particles is assumed, carrying a fraction $\epsilon_e$ of the local internal energy density and with power law slope $-p$ (not to be confused with pressure $p$). The fraction of available electrons that is accelerated is given by $\xi_N$. A further fraction $\epsilon_B$ of the local internal energy density resides in the shock-generated magnetic field. The effect of electron cooling is included using a global estimate for the electron cooling time $t_c$, by equating it to the lab frame time since the explosion. The spectral shape of the absorption coefficient $\alpha_\nu$ for s.s.a. is also given by sharply connected power laws and the effect of electron cooling on $\alpha_\nu$ is ignored (in any case, the error thus introduced is neglible compared to the small error from using a global rather than a local estimate for electron cooling. Global and local electron cooling are compared in \citealt{vanEerten2010offaxis}). Mathematical expressions for the emission and absorption coefficients can be found in \citealt{vanEerten2012boxfit}.

After all rays have emerged from the blast wave and the observed flux is calculated by integrating over the rays, the observed spectrum will again consist of a set of power laws, now smoothly connected due to the different break frequencies at different contributing parts of the fluid. S.s.a. manifests itself as an additional break in the spectrum, occurring typically at radio wavelengths. An example spectrum, calculated from our Lorentz-boosted simulation for $\theta_0 = 0.05$ rad and ISM environment is provided by Fig. \ref{powerlaws_figure}, and reveals how the full syncrotron spectrum can be reproduced even at $10^{-2}$ days. A comparison between an X-ray light curve from the boosted frame and from \textsc{boxfit} \citep{vanEerten2012boxfit} is given by the bottom plot of Fig. \ref{resolution_figure}. Note that the light curve is produced by the integrated emission from the entire blast wave and as a result, the resolution difference in the dynamics as shown in the top panel of \ref{resolution_figure} does not directly reflect the discrepancy in the observed emission. Whether a lower resolution for the blast wave dynamics leads to an overestimate or an underestimate of the flux depends on the spectral regime that is observed. Although Fig. \ref{powerlaws_figure} includes s.s.a., we will focus in this work on optical and X-ray frequencies (where the jet break is typically observed) and postpone a detailed treatment of the s.s.a. break to future work.

\subsection{Resolution}

All light curves were calculated for $p = 2.5$, which is sufficient to derive light curves for arbitrary $p > 2$ value, as explained below in section \ref{criticals_section}. Lacking an upper cut-off to the accelerated particle energy distribution, our radiation model is invalid for $p \le 2$ because then the integral for the total accelerated particle energy diverges. The following settings were used to compute numerically converged synthetic light curves from the boosted frame simulations. 3000 simulation snapshots were probed for computing on-axis light curves, 300 snapshots were probed for computing off-axis light curves. A matrix of rays was used consisting of 1500 rays logarithmically spaced in the radial direction and 100 evenly spaced in the angular direction (or 1 in the angular direction, for on-axis observations). These directions refer to coordinates on the plane perpendicular to the observer (in the lab frame, see appendix \ref{boosted_frame_appendix}), and the inner and outer boundaries of this plane are $10^{12}$ and $10^{18}$ cm respectively. Each light curve has 150 data points between observer times of $10^{-4}$ days and $10^2$ days, although only 85 (60) data points between $0.01$ (0.1) and 26 days were used for analysis in order to ensure complete coverage of the observer times by the emission times. Light curves were calculated for half-opening angles $\theta_0 = 0.05$, $0.1$ and $0.2$ rad and observer angles $\theta_{obs}$ that were a fraction 0, 0.2, 0.4, 0.6, 0.8 or 1 of $\theta_0$.

\section{Peak flux and break frequencies}
\label{criticals_section}

\subsection{Theory}

In section \ref{initial_conditions_subsection} we demonstrated the scale-invariance of the jet dynamics between different jet energies and circumburst densities. In \cite{vanEerten2012scalings} we demonstrated that similar scalings apply in the asymptotic regimes of the observed spectrum. Although additional constants are introduced when calculating synchrotron radiation, such as the electron mass $m_e$, these can be identified and isolated in the flux equations in a given spectral regime and the resulting scale invariance is again a result of dimensional analysis.

Invariance of the fluxes in each asymptotic spectral regime is equivalent to scale invariance of the critical quantities that determine the shape of the spectrum: peak flux $F_{peak}$, synchrotron break frequency $\nu_m$, cooling break frequency $\nu_c$ and s.s.a. break frequency $\nu_a$. The shapes of the spectral transitions are not scale-invariant, but can be modeled using a smooth connection between power laws (see \citealt{Granot2002, vanEerten2009BMscalingcoefficients, Leventis2012}). The flux for a given observation can be calculated from the characteristic scale-invariant evolution of the critical quantities plus a description of the spectral transitions (which can also be a simple sharp power law approximation), and from three sets of parameters defining the observation and model: the observer parameters $z$ (redshift), $d_L$ (luminosity distance), $\theta_{obs}$ (observer angle), $t_{obs}$ (time) and $\nu$ (frequency); the explosion parameters $k$, $A$, $E_{iso}$, $\theta_0$; the radiation 
parameters $p$, $\epsilon_B$, $\epsilon_e$, $\xi_N$. The dependency of the flux on $d_L$ and $z$ is straightforward, with $F_\nu \propto d_L^{-2}$ and flux, frequency and time depending on $z$ in the standard manner. Different values for $\theta_{obs}$, $\theta_0$ and $k$ lead to different evolution of the characteristic quantities $\nu_a$, $\nu_m$, $\nu_c$ and $F_{peak}$. Scale invariance takes care of $E_{iso}$ and $A$, while the dependency of the characteristic quantities on $\epsilon_B$, $\epsilon_e$, $\xi_N$ remains unchanged throughout the evolution of the decollimating blast wave and can be determined analytically for general $p$ and $k$. The dependency of the characteristic quantities on $p$ is constant in time and known analytically, meaning that once the evolution for a given $p$ value is known, their evolution for any $p$ value can be trivially obtained. Different spectral orderings of the break frequencies $\nu_a$, $\nu_m$, $\nu_c$ lead to different time evolutions.

\begin{table}
\centering
\begin{tabular}{|l|}
\hline
\\
\normalsize
$\displaystyle \kappa \equiv \left( \frac{E_{iso}}{10^{53} \textrm{ erg}} \right)$\\[15pt]
\normalsize
$ \displaystyle \lambda \equiv \left( \frac{n_{0, ref}}{1 \textrm{ cm}^{-3}} \right)$ \\[15pt]
\normalsize
$ \displaystyle \tau \equiv (\lambda / \kappa)^{1/(3-k)} t_{obs} / (1+z)$ \\[15pt]
\hline
\\
\normalsize
$ \displaystyle F_{peak} = \frac{(1+z)}{d^2_{28}}  \frac{p-1}{3p-1}  \epsilon_e^0  \epsilon_B^{1/2}  \xi_N^1 \kappa^{\frac{3(2-k)}{2(3-k)}}  \lambda^{\frac{3}{2(3-k)}}  \mathfrak{f}_{peak} (\tau ; \theta_0, \theta_{obs}, k)$ \\[15pt]
\normalsize
$ \displaystyle \nu_{m} = (1+z)^{-1}  \left( \frac{p-2}{p-1} \right)^2  \epsilon_e^2  \epsilon_B^{1/2}  \xi_N^{-2}  \kappa^{\frac{-k}{2(3-k)}}  \lambda^{\frac{3}{2(3-k)}}  \mathfrak{f}_{m} (\tau ; \theta_0, \theta_{obs}, k)$ \\[15pt]
\normalsize
$ \displaystyle \nu_{c} = (1+z)^{-1}  \epsilon_e^0  \epsilon_B^{-3/2}  \xi_N^0  \kappa^{\frac{3k-4}{2(3-k)}}  \lambda^{\frac{-5}{2(3-k)}}  \mathfrak{f}_{c} (\tau ; \theta_0, \theta_{obs}, k)$ \\[15pt]
\hline
\\
\normalsize
$ \displaystyle F_{peak} = \frac{(1+z)}{d^2_{28}} \frac{p-1}{3p-1}  \epsilon_e^0  \epsilon_B^{1/2}  \xi_N^1  \kappa  \lambda^{1/2}  \mathfrak{f}_{peak} (\tau ; \theta_0, \theta_{obs}, k = 0)$ \\[15pt]
\normalsize
$ \displaystyle \nu_{m} = (1+z)^{-1}  \left( \frac{p-2}{p-1} \right)^2  \epsilon_e^2  \epsilon_B^{1/2}  \xi_N^{-2}  \kappa^0  \lambda^{1/2}  \mathfrak{f}_{m} (\tau ; \theta_0, \theta_{obs}, k = 0)$ \\[15pt]
\normalsize
$ \displaystyle \nu_{c} = (1+z)^{-1}  \epsilon_e^0  \epsilon_B^{-3/2}  \xi_N^0  \kappa^{-2/3}  \lambda^{-5/6}  \mathfrak{f}_{c} (\tau ; \theta_0, \theta_{obs}, k = 0)$ \\[15pt]
\hline
\\
\normalsize
$ \displaystyle F_{peak} = \frac{(1+z)}{d^2_{28}}  \frac{p-1}{3p-1}  \epsilon_e^0  \epsilon_B^{1/2}  \xi_N^1  \kappa^0  \lambda^{3/2}  \mathfrak{f}_{peak} (\tau ; \theta_0, \theta_{obs}, k = 2)$ \\[15pt]
\normalsize
$ \displaystyle \nu_{m} = (1+z)^{-1}  \left( \frac{p-2}{p-1}  \right)^2  \epsilon_e^2  \epsilon_B^{1/2}  \xi_N^{-2}  \kappa^{-1}  \lambda^{3/2}  \mathfrak{f}_{m} (\tau ; \theta_0, \theta_{obs}, k = 2)$ \\[15pt]
\normalsize
$ \displaystyle \nu_{c} = (1+z)^{-1} \epsilon_e^0 \epsilon_B^{-3/2} \xi_N^0 \kappa^{1} \lambda^{-5/2} \mathfrak{f}_{c} (\tau ; \theta_0, \theta_{obs}, k = 2)$ \\[15pt]
\hline
\end{tabular}
\caption{Concise equations for the characteristic quantities. Top panel shows energy and density scale factors $\kappa$ and $\lambda$ and scaled time $\tau$. The following panels show equations for general $k$, $k = 0$ (ISM) and $k = 2$ (stellar wind) from top to bottom.}
\label{characteristics_table}
\end{table}

Table \ref{characteristics_table} summarizes these properties of the light curves. Here the general $k$ case is shown as well as the ISM and stellar wind cases separately. $d_{28}$ is the luminosity distance $d_L$ in units of $10^{28}$ cm. The functions $\mathfrak{f}_{peak} (\tau ; \theta_0, \theta_{obs}, k)$, $\mathfrak{f}_{m} (\tau ; \theta_0, \theta_{obs}, k)$ and $\mathfrak{f}_{c} (\tau ; \theta_0, \theta_{obs}, k)$ denote the scale-invariant time evolution of the characteristic quantities as they are determined numerically from analyzing light curves computed from the boosted frame simulations for each spectral regime. These functions can be scaled from their baseline values to arbitrary explosion energy and circumburst density by plugging in the scaled values for $\kappa$, $\lambda$ and $\tau$ from the top section of the table into the equations in the lower sections of the table. Since their dependency on the radiation parameters $\epsilon_B$, $\epsilon_e$, $\xi_N$ and $p$ is known and constant in time, these terms have been made explicit in the table. Left implicit is the fact that different spectral orderings lead to different evolution curves, which will affect the characteristic scale-invariant functions $\mathfrak{f}$ but not their pre-factors. In the remainder of this paper we will discuss the \emph{slow cooling} case (which is the one usually observed in practice), where $\nu_m < \nu_c$. The observed flux follows from the characteristic quantities according to
\begin{eqnarray}
 F_D & = & F_{peak} \left( \frac{\nu}{\nu_m} \right)^{1/3}, \qquad \nu < \nu_m < \nu_c, \nonumber \\
 F_G & = & F_{peak} \left( \frac{\nu}{\nu_m} \right)^{(1-p)/2}, \qquad \nu_m < \nu < \nu_c, \nonumber \\
 F_H & = & F_{peak} \left( \frac{\nu_c}{\nu_m} \right)^{(1-p)/2} \left( \frac{\nu}{\nu_c} \right)^{-p/2}, \qquad \nu_m < \nu_c < \nu,
\label{flux_equations}
\end{eqnarray}
where the labels $D$, $G$, $H$ have been chosen to match the notation from \cite{Granot2002}. Note that, even though the time evolution of the characteristic quantities does not depend on $p$, equations \ref{flux_equations} imply that the time evolution of $F_G$ and $F_H$ do.

\subsection{Numerical results}

\begin{figure}
 \centering
  \includegraphics[width=\columnwidth]{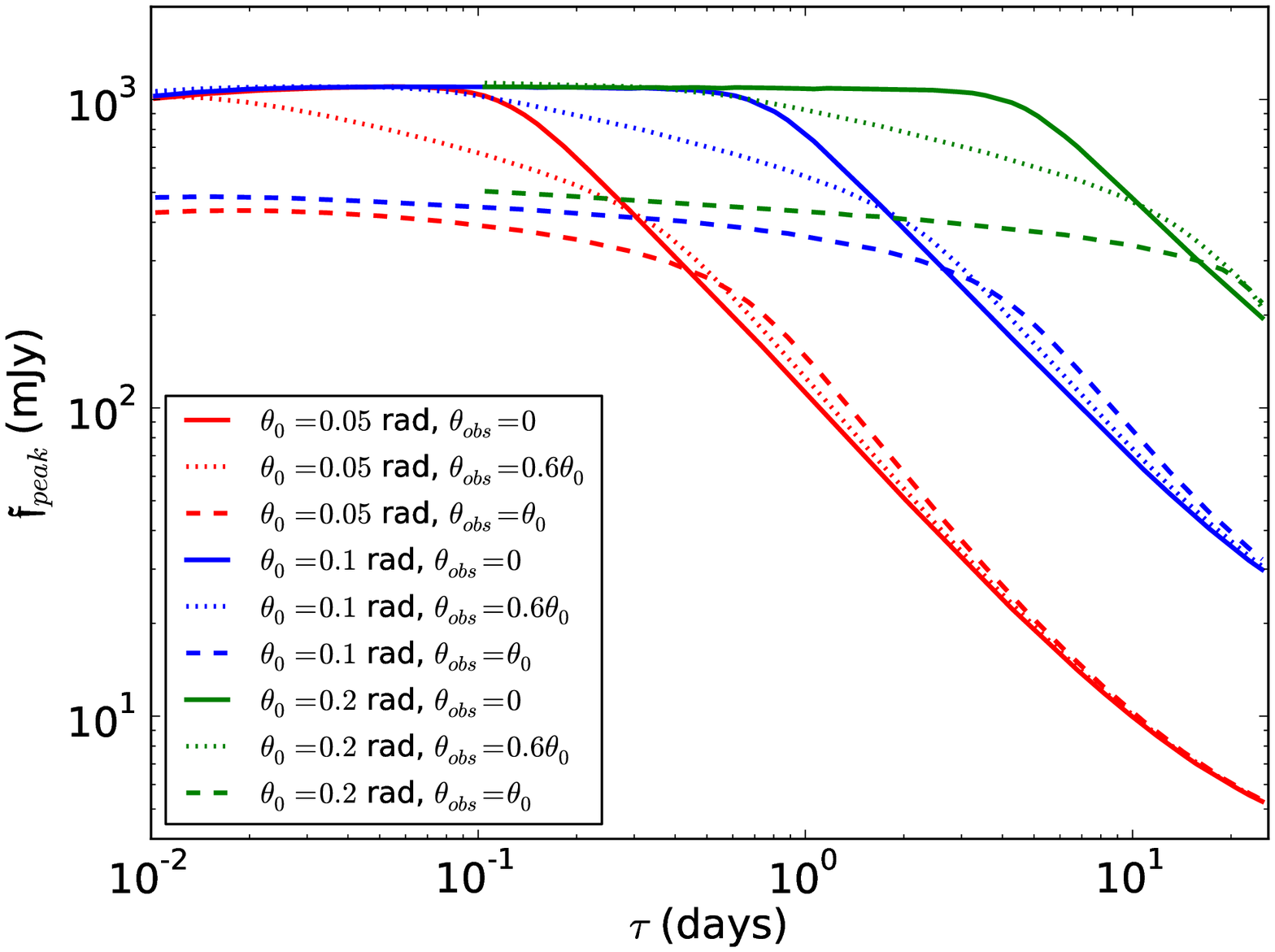}
  \includegraphics[width=\columnwidth]{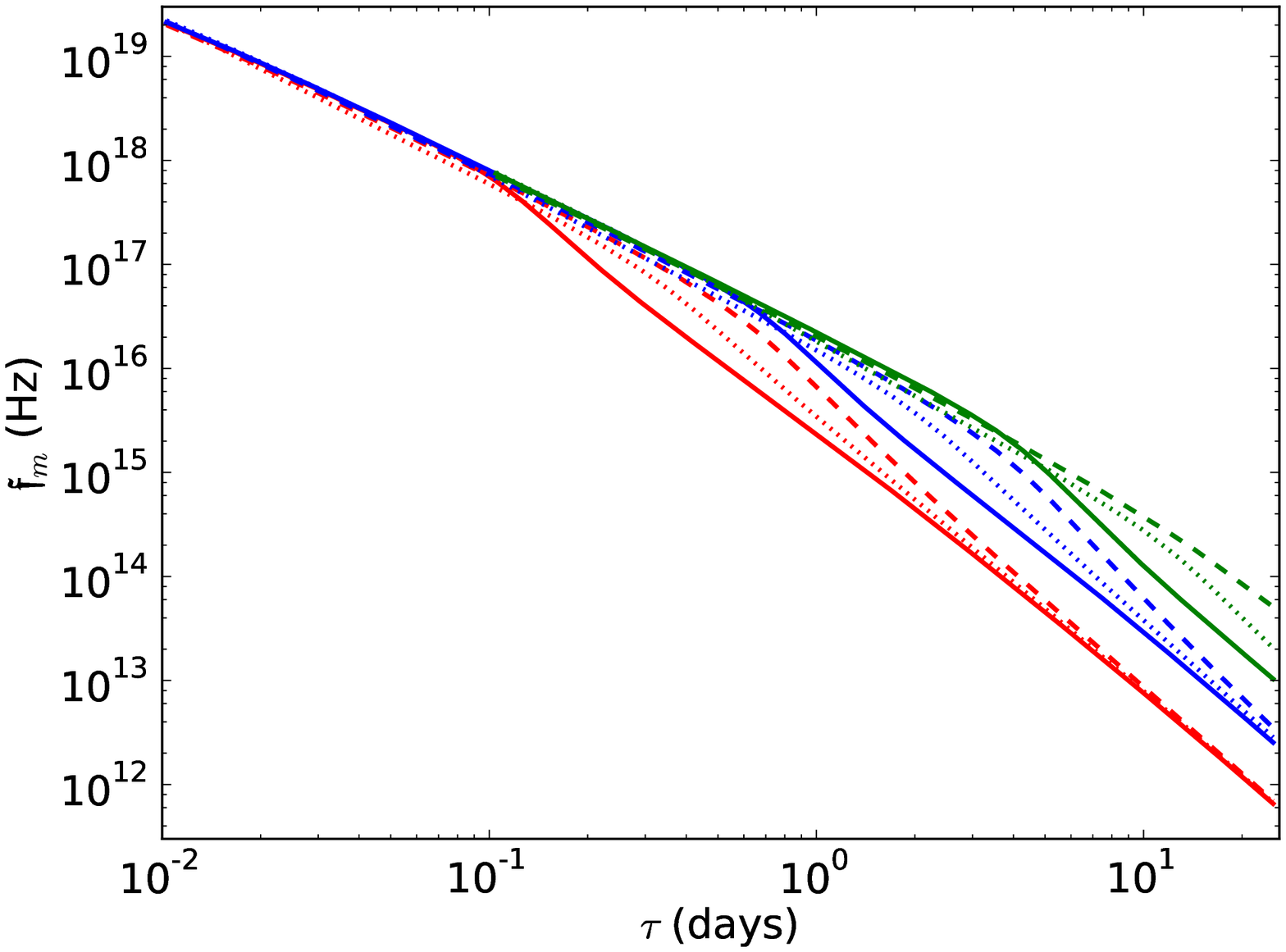}
  \includegraphics[width=\columnwidth]{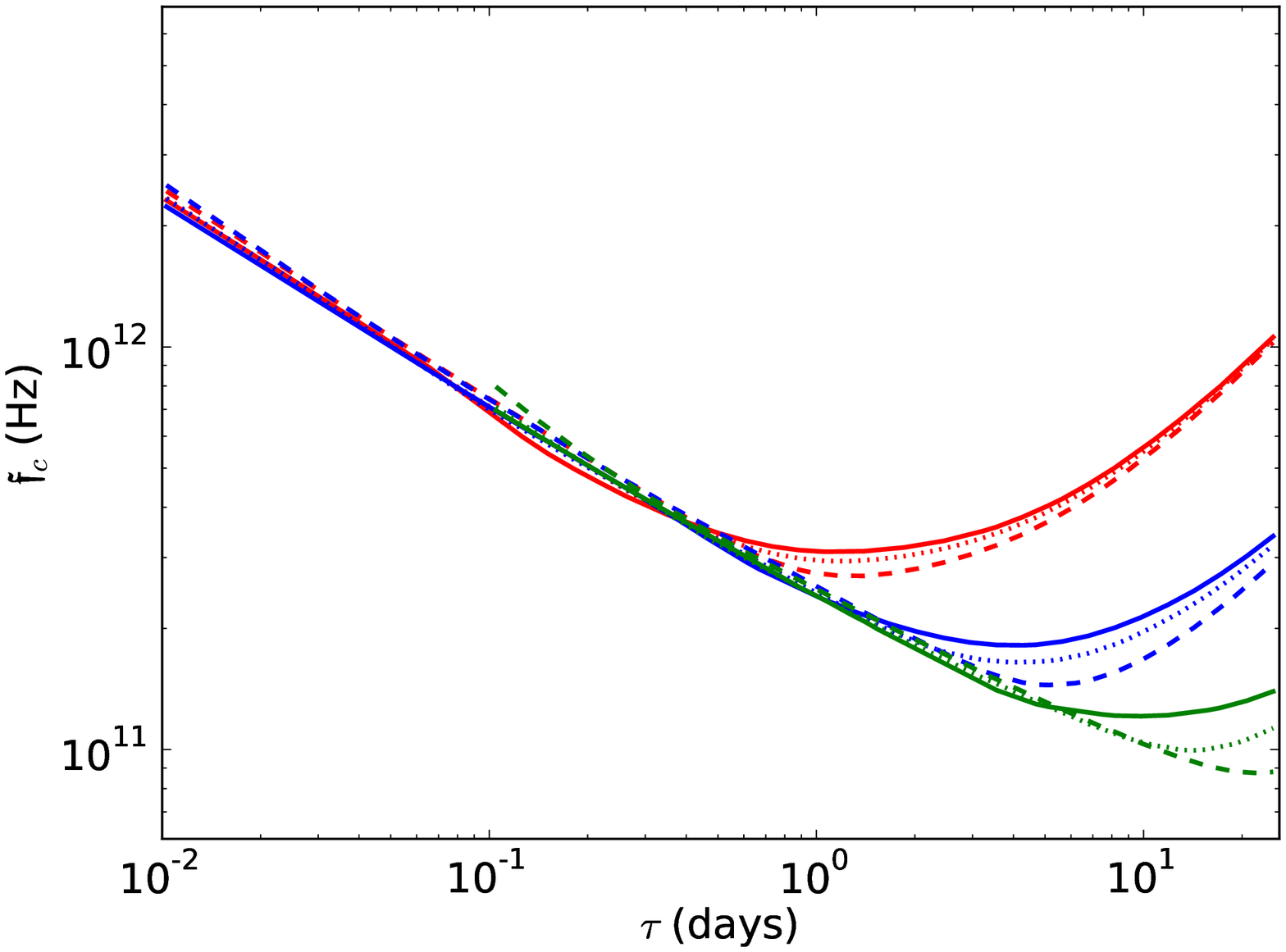}
  \caption{Time evolution of characteristic quantities, for different jet opening angles and observer angles. Top to bottom: $\mathfrak{f}_{peak}$, $\mathfrak{f}_{m}$, $\mathfrak{f}_{c}$. They describe the time evolution of $F_{peak}$, $\nu_m$, $\nu_c$ respectively according to the equations in table \ref{characteristics_table}. The legend in the top plot applies to all plots.}
\label{characteristics_plot}
\end{figure}

\begin{figure}
 \centering
  \includegraphics[width=\columnwidth]{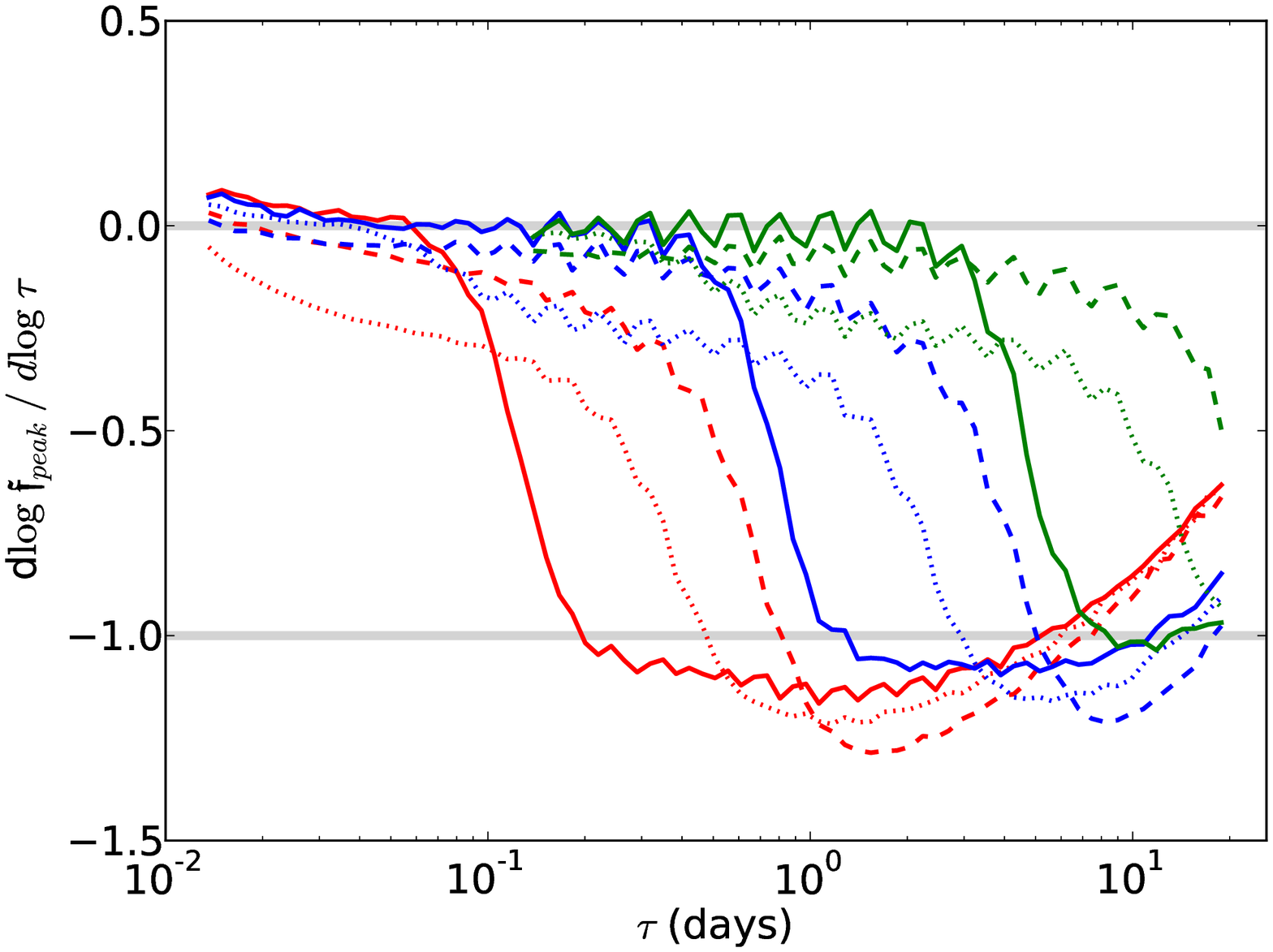}
  \includegraphics[width=\columnwidth]{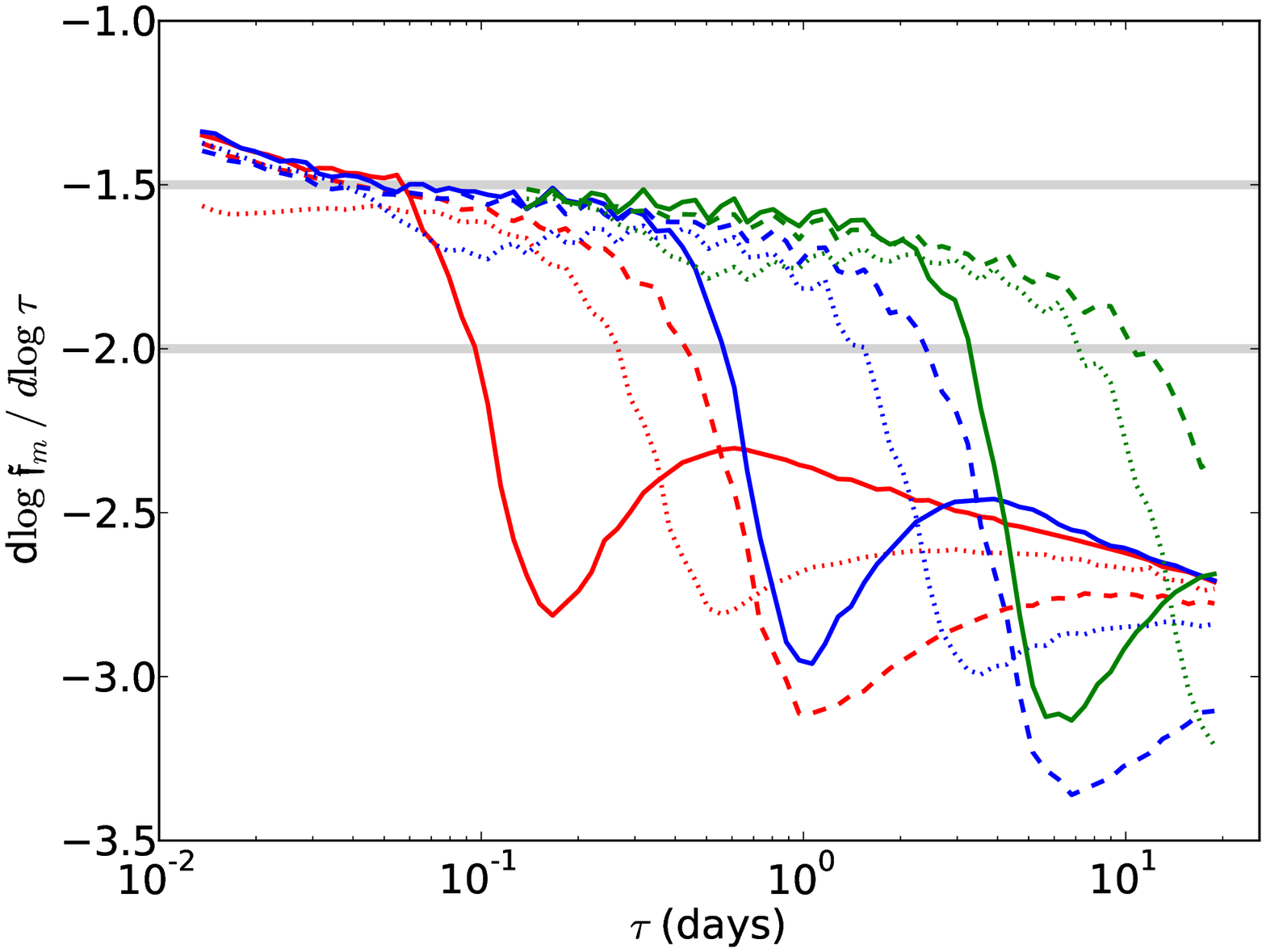}
  \includegraphics[width=\columnwidth]{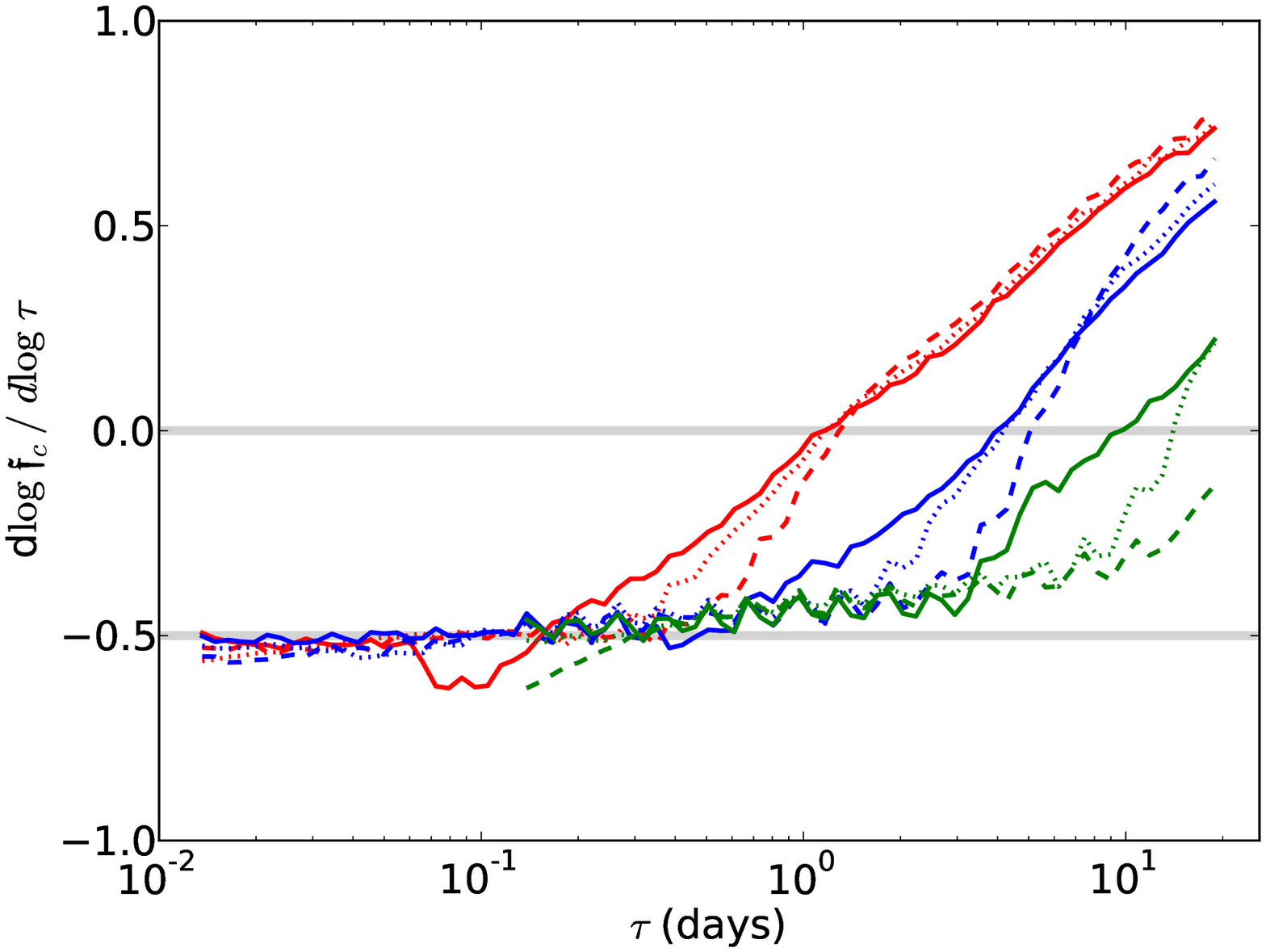}
  \caption{Time evolution of power law slopes of the characteristic quantities, for different jet opening angles and observer angles. Top to bottom: slopes for $\mathfrak{f}_{peak}$, $\mathfrak{f}_{m}$, $\mathfrak{f}_{c}$. The legend in the top plot of Fig. \ref{characteristics_plot} applies to all these plots as well. The constant grey lines indicate the expected slopes for light curves for an on-axis observer from the pre-break BM solution and assuming a fast spreading jet post-break.}
\label{characteristics_slopes_plot}
\end{figure}

In Fig. \ref{characteristics_plot} we plot the time evolution of the characteristic quantities for the three jet opening angles $\theta_0 = 0.05$, $0.1$, $0.2$ rad. and for observer angles $\theta_{obs} = 0$, $0.6 \times \theta_0$, $\theta_0$. Fig. \ref{characteristics_slopes_plot} shows the evolution of the spectral slope for these same angles. The time evolutions were calculated from three light curves per $\theta_0$, $\theta_{obs}$ combination: one for each asymptotic spectral regime separated by $\nu_m$ and $\nu_c$ in the slow cooling case. For these light curves we used $\epsilon_B = 10^{-5}$, $\epsilon_e = 10^{-5}$, $\xi_N = 1$, $p = 2.5$ and frequencies $10^{-20}$, $10^{10}$, $10^{40}$ Hz. These values (especially the frequencies) were not physically motivated but rather chosen such that they ensure all light curves were calculated well into the asymptotic limits of the spectral regimes and not impacted by the smoothness of the spectral transitions between regimes. Moving the outer frequencies closer in but still in their asymptotic regions throughout the evolution of the emission for $\epsilon_B = 10^{-5}$, $\epsilon_e = 10^{-5}$, $\xi_N = 1$, $p = 2.5$, e.g. to $10^{-5}$ and $10^{25}$ was found to have no impact on the result. The characteristic quantities were subsequently obtained by inverting equations \ref{flux_equations}. Since the plots show $\mathfrak{f}_{peak}$, $\mathfrak{f}_m$ and $\mathfrak{f}_c$, the curves are independent of $\epsilon_e$, $\epsilon_B$ and $p$. 

Figs. \ref{characteristics_plot} and \ref{characteristics_slopes_plot} reveal that the time evolution for the characteristic quantities strongly depends on both jet and observer angle. A difference between pre- and post-break values sets in immediately following the jet break time $t_j$, which for the current ISM simulations and on-axis observers is found to lie around
\begin{equation}
 t_j \approx (0.6\pm0.1) (1+z) ( \kappa / \lambda )^{1/3} (\theta_0 / 0.1)^{8/3} \textrm{ days,}
\end{equation}
consistent with the earlier reported numerical results from \cite{vanEerten2010offaxis}. The jet break time will be discussed in more detail in section \ref{lightcurves_section} below. In Fig. \ref{characteristics_slopes_plot}, both pre- and post-break theoretically expected slopes are also plotted. The pre-break slopes match the theoretical predictions well, but the post-break slopes differ substantially from theoretical predictions (see \citealt{Sari1999, Rhoads1999}). 

Partly this discrepancy between theory and numerical practice is a consequence of the fact that the spreading of blast waves in simulations is not an exponential process even for $\theta_0 = 0.05$ rad. \citep{vanEerten2012observationalimplications, MacFadyen2013}. The fact that $\nu_m$ is far more strongly impacted by jet spreading than theoretically expected and even more so than $F_{peak}$ can be understood as follows. Considering intensities rather than surface integrated flux (i.e. $I_{peak}$ rather than $F_{peak}$), which leaves the angular dependency explicit, we have $I_{peak} \propto (1 - \beta \mu)^{-3}$. Here $\beta$ is the outflow velocity in terms of $c$, $\mu$ the cosine of the angle between flow and observer direction. The expression includes the effect of departure time difference between emission from front and back of the blast wave as well as the Lorentz transform of the emission coefficient (see also the appendix of \citealt{vanEerten2010offaxis}). On the other hand, for $\nu_{m, I}$, which we define as the contribution to $\nu_m$ along a single beam, we have $\nu_{m,I} \propto (1- \beta \mu)^{-1}$. While both $I_{peak}$ and $\nu_{m,I}$ are beamed it therefore follows that the beaming effect is far stronger for $I_{peak}$, such that $F_{peak}$ is then less sensitive for the behavior of the flow near the edges than $\nu_m$. Although in theory this effect could be compensated for by a strong dependence of $I_{peak}$ and $\nu_{m,I}$ on emission time (since edge emission arriving at the same time departs earlier than emission along the axis to the observer), it turns out in practice that this only strengthens the sensitivity of $\nu_{m,I}$ to emission angle compared to the angle dependence of $I_{peak}$: for the BM solution, the scalings are $\nu_{m,I} \propto t^{-3} (1 - \beta \mu)^{-1}$ and $I_{peak} \propto t^4 (1 - \beta \mu)^{-3}$.

\begin{figure}
 \centering
  \includegraphics[width=\columnwidth]{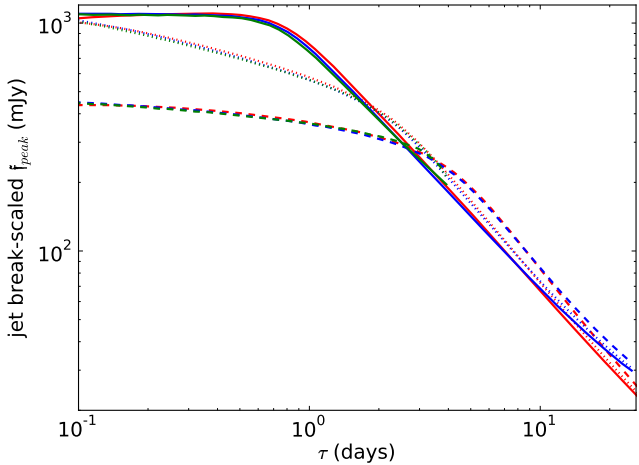}
  \caption{Scaled evolution of the peak flux function $\mathfrak{f}_{peak}$, where the curves for $\theta_0 = 0.05$ rad. and $\theta_0 = 0.2$ rad. have been scaled in time towards the $\theta_0 = 0.1$ rad. result using $t' = (\theta_0 / 0.1)^{8/3}$. As in Figs. \ref{characteristics_plot} and \ref{characteristics_slopes_plot}, solid lines refer to $\theta_{obs} = 0$, dotted lines to $\theta_{obs} = 0.6 \theta_0$ and dashed lines to $\theta_{obs} = \theta_0$.}
\label{jetbreak_scaled_plot}
\end{figure}

It would be a strong indication of self-similarity between jet opening angles and of great practical significance if the evolution functions were to scale in a straightforward manner between opening angles. Any such scaling should incorporate the $\theta_0$ dependency of jet break time $t_j$. When we take that as our starting point, scale time according to $t' = t (\theta'_0 / \theta_0)^{8/3}$ and the characteristic functions according to $\mathfrak{f}'(t') = (\theta'_0 / \theta_0)^{-\alpha} \mathfrak{f}(t)$, where $\alpha$ is the power law time-dependence in the pre-break BM regime, this yields evolution curves for $\mathfrak{f}_{peak}$ that numerically match quite well initially between different jet opening angles, even for off-axis observer angles, as illustrated in Fig. \ref{jetbreak_scaled_plot}. These $\theta_0$-scalings however are not exact and a similar mapping for $\mathfrak{f}_m$ or $\mathfrak{f}_c$ fails to produce much numerical overlap. This can be seen from the power law slope plots in Fig. \ref{characteristics_slopes_plot}, since the scalings represent horizontal shifts of the curves in these plots. Essentially, this lack of straightforward scalability reflects the fact that the post-break behavior is determined by more characteristic timescales than $t_j$ alone, such as the transition time to non-relativistic flow and the transition time to quasi-spherical flow, and that  these timescales will impact the trans-relativistic stage of fluid flow that generates the observed post-break light curves.

\subsection{The cooling break}

Figs. \ref{characteristics_plot} and \ref{characteristics_slopes_plot} illustrate that for a given characteristic function different extremal values are reached for different observer and jet angles. The consequence of this on the jet break as measured from observations will be discussed in section \ref{lightcurves_section} below, and we limit ourselves here to highlighting the behavior of the cooling break. In an earlier work \citep{vanEerten2012scalings} we showed how simulations in a fixed frame (and thus of lower resolution) indicated a steepening (for $\theta_{obs} = 0$) of the temporal evolution of the cooling break immediately following the jet break. However, the current boosted frame simulations \emph{reveal no post-break $\nu_c$ steepening towards stronger decay}. Specifically, the $\nu_c$ light curves for $\theta_0 = 0.2$ rad., the same angle as plotted in Fig. 3 of \cite{vanEerten2012scalings}, show only a turnover towards positive temporal slope following the jet break, as can be seen in the bottom panel of Fig. \ref{characteristics_slopes_plot} of the current paper. At the same time, the on-axis curve for $\theta_{obs} = 0.05$ rad. shown in the same figure, does show a (slight) post-break steepening of the temporal power law slope of $\nu_c$. 

\begin{figure}
 \centering
  \includegraphics[width=\columnwidth]{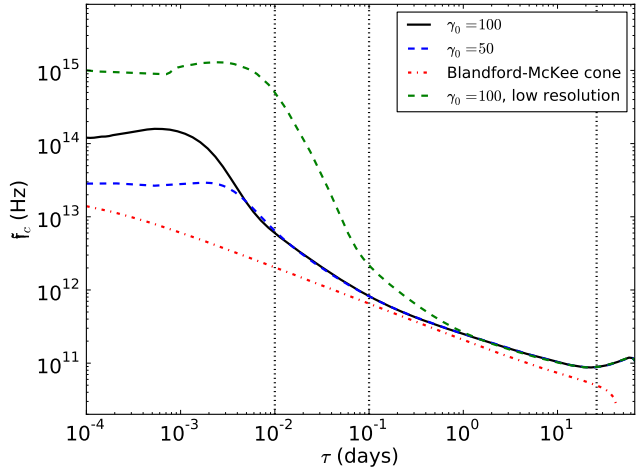}
  \caption{A comparison of the time evolution of four computations of $\nu_c$ for $\theta_0$ = 0.2 rad. and $\theta_{obs} = \theta_0$ (i.e. an on-edge observer). The vertical lines indicate the timespans used for analysis.}
\label{nuc_comparison_plot}
\end{figure}

What this indicates is that the earlier reported steepening for $\theta_0 = 0.2$ rad. and the current smaller steepening for $\theta_0 = 0.05$ rad. are numerical in origin, and sensitive to the initial conditions of the blast wave. Above the cooling break, the observed flux is dominated by emission from the edges of the jet (i.e. the observed image is `limb-brightened'), and therefore the cooling break $\nu_c$ is the most sensitive to deviations from purely radial flow at the edges of an initially conically truncated spherical BM outflow. The smaller the jet opening angle, the larger even small resolution-induced deviations are relative to $\theta_0$. The dynamics of narrow and wide jets will be discussed separately in more detail in \cite{MacFadyen2013}.
 
The effect of early time flow at the jet edges on the light curve naturally becomes more severe the closer the observer angle moves towards the edge of the jet. In Fig. \ref{nuc_comparison_plot}, we show that even the pre-break behavior for $\nu_c$ from 2D simulations differs strongly from that from analytically calculated conical outflow. The black solid line and blue dashed line show $\nu_c$ results for simulations starting at $\gamma_0 = 100$ and $\gamma_0 = 50$ respectively, the red dash-dotted line shows the evolution of $\nu_c$ based on conical outflow following the BM solution, while the green dashed line shows a four times lower resolution simulation. Around the leftmost vertical line, the $\nu_c$ curves for both normal resolution simulations have merged, while both simulation curves still differ strongly from the BM solution. It follows that the difference between 2D simulated and radial analytical flow can not be attributed to a lack of early time coverage of the observed signal by an incomplete range of emission times. Nor can this difference be attributed to the difference in starting times (and hence the extent to which $\gamma_0 \ll 1 / \theta_0$. Both effects are clearly visible in Fig \ref{nuc_comparison_plot} and lie well to the left of the left vertical line at $10^{-2}$ days. In view of this resolution issue, when analysing light curves for $\theta_0 = 0.2$ rad we will start from 0.1 days (rather than 0.01 days), and the characteristic evolution curves for this initial jet opening angle in Figs \ref{characteristics_plot} and \ref{characteristics_slopes_plot} have been truncated at this value of $\tau$. Note that for most observer angles, this effect is less severe and the parameters of Fig \ref{nuc_comparison_plot} were chosen to reflect a worst-case scenario.

An additional conclusion that can be drawn from the severe resolution dependence of the off-axis observed $\nu_c$ evolution for $\theta_0 = 0.2$ rad., even well into times that are easily observable by instruments such as \emph{Swift}, is that if small numerical resolution-induced deviations from BM-type flow will have a large effect on $\nu_c$, the same will hold for minor \emph{physical} deviations. This renders relevant the question to what extent deviations from the expected BM-based time evolution of $\nu_c$ can be driven by the dynamics of the outflow. On the other hand, although an actual measurement of the evolution of $\nu_c$ has been performed by \cite{Filgas2011}, the temporal slope of -1.2 that these authors find is steeper than the high-resolution simulation $\nu_c$ slope in Fig. \ref{characteristics_slopes_plot} at any time, and their thesis is that the steep decline in GRB 091127 can be attributed to changes in the radiative process (via a time dependency of $\epsilon_B$) rather than outflow 
dynamics.

\section{Light curves and jet breaks}
\label{lightcurves_section}

Once the time evolution of the characteristic functions $\mathfrak{f}_{peak}$, $\mathfrak{f}_c$ and $\mathfrak{f}_c$ is known, they can be used to quickly calculate light curves for arbitrary $p$. In order to study the shape of the jet break we have done so for $p$ values of $2.01$ and $2.1$, $2.2$, $\ldots$, $3.0$ and the spectral regimes $\nu_a < \nu_m < \nu < \nu_c$ (typically applicable to optical data) and $\nu_a < \nu_m < \nu_c < \nu$ (``X-rays''). Various functions have been used in the literature to fit jet breaks in optical and X-ray light curve data, such as sharp power laws (e.g. \citealt{Racusin2009, Evans2009}), smoothly connected power laws (e.g. \citealt{Beuermann1999}) or power law transitions where the turnover includes an exponential term (e.g. \citealt{Harrison1999}). A limitation common to all these fit functions is the assumption of a single power law regime after the jet break. Although simulation-based light curves show that in reality this should not be expected to be the case (as can be seen from the post-break evolution of the peak flux and break frequencies, Figs. \ref{characteristics_plot} and \ref{characteristics_slopes_plot}), it therefore makes practical sense to explore the implications of our simulation results for the interpretation and applicability of broken power law fit functions.

\begin{table*}
\centering
\begin{tabular}{|ll|l|llllll|ll|}
\hline
$\theta_0$ (rad) & $\theta_{obs}$ & fit & $\alpha_0$ & $\alpha_1$ & $< \tau_b >$ & $^{10}\log \bar{C}$ & $\sigma$ & $<\tau_{0.9} / \tau_b>$ & $\chi^2 / \chi^2_{PL}$ & $\chi^2$, red.$^\dagger$ \\
\hline
0.05 & 0 & PL & $0.76 -0.73 p$ & $0.20 -1.24 p$ & 0.10 & $1.37 + 1.64 p$ & & &  1 & 0.20 \\
 & & sB & $0.75 -0.71 p$ & $0.19 -1.24 p$ & 0.10 & $1.34 + 1.68 p$ & $6.06 -0.91 p$ & 1.4 &   0.68 & 0.14 \\
 & & sH & $1.01 -0.74 p$ & $0.18 -1.23 p$ & 0.08 & $1.57 + 1.66 p$ & & &  1.00 & 0.20 \\
 & $0.2 \theta_0$ & PL & $0.69 -0.74 p$ & $0.19 -1.24 p$ & 0.12 & $1.34 + 1.60 p$ & & &  1 & 0.54 \\
 & & sB & $0.71 -0.69 p$ & $0.20 -1.25 p$ & 0.10 & $1.33 + 1.68 p$ & $3.65 -0.70 p$ & 1.8 &  0.34 & 0.18 \\
 & & sH & $1.02 -0.78 p$ & $0.19 -1.24 p$ & 0.09 & $1.66 + 1.58 p$ & & &  0.28 & 0.15 \\
 & $0.4 \theta_0$ & PL & $0.67 -0.82 p$ & $0.23 -1.28 p$ & 0.17 & $1.54 + 1.35 p$ & & &  1 & 0.99 \\
 &  & sB & $0.72 -0.77 p$ & $0.23 -1.29 p$ & 0.15 & $1.53 + 1.45 p$ & $2.92 -0.51 p$ & 2.1 & 0.58 & 0.57\\
 &  & sH & $0.94 -0.85 p$ & $0.22 -1.28 p$ & 0.14 & $1.86 + 1.35 p$ & & & 0.46 & 0.46  \\
 & $0.6 \theta_0$ & PL & $0.56 -0.80 p$ & $0.26 -1.32 p$ & 0.25 & $1.52 + 1.22 p$ & & &  1 & 0.74 \\
 & & sB & $0.58 -0.79 p$ & $0.26 -1.32 p$ & 0.24 & $1.52 + 1.24 p$ & $5.63 -0.92 p$ & 1.5 & 0.85 & 0.64 \\
 & & sH & $0.78 -0.83 p$ & $0.24 -1.32 p$ & 0.21 & $1.83 + 1.19 p$ & &  & 0.87 & 0.65 \\
 & $0.8 \theta_0$ & PL & $0.56 -0.77 p$ & $0.29 -1.37 p$ & 0.33 & $1.44 + 1.16 p$ & & & 1 & 1.07 \\
 & & sB & $0.56 -0.75 p$ & $0.29 -1.37 p$ & 0.32 & $1.40 + 1.20 p$ & $5.72 -1.09 p$ & 1.5 & 0.86 & 0.93 \\
 & & sH & $0.74 -0.79 p$ & $0.27 -1.37 p$ & 0.29 & $1.69 + 1.14 p$ & &  & 0.84 & 089 \\
 & $\theta_0$ & PL & $0.66 -0.77 p$ & $0.31 -1.41 p$ & 0.43 & $1.34 + 1.12 p$ & & &  1 & 1.48  \\
 & & sB & $0.68 -0.76 p$ & $0.33 -1.42 p$ & 0.41 & $1.39 + 1.13 p$ & $4.26 -0.74 p$ & 1.6 & 0.82 & 1.23 \\
 & & sH & $0.82 -0.79 p$ & $0.31 -1.42 p$ & 0.38 & $1.62 + 1.09 p$ & &  & 0.75 & 1.11 \\
\hline
 0.1 & 0 & PL & $0.74 -0.75 p$ & $0.26 -1.29 p$ & 0.59 & $1.90 + 1.08 p$ & & & 1 & 0.23 \\
 & & sB & $0.75 -0.75 p$ & $0.25 -1.29 p$ & 0.58 & $1.91 + 1.08 p$ & $5.35 -0.77 p$ & 1.4 & 0.61 & 0.14 \\
 & & sH & $0.85 -0.76 p$ & $0.21 -1.28 p$ & 0.54 & $2.06 + 1.07 p$ & &  & 1.11 & 0.24 \\
 & $0.2 \theta_0$ & PL & $0.73 -0.76 p$ & $0.26 -1.29 p$ & 0.64 & $1.92 + 1.03 p$ & & &  1 & 0.66 \\
 & & sB & $0.74 -0.74 p$ & $0.26 -1.31 p$ & 0.63 & $1.93 + 1.07 p$ & $2.96 -0.52 p$ & 2.0 & 0.17 & 0.12 \\
 & & sH & $0.85 -0.77 p$ & $0.22 -1.29 p$ & 0.59 & $2.11 + 1.02 p$ & & & 0.10 & 0.06 \\
 & $0.4 \theta_0$ & PL & $0.73 -0.80 p$ & $0.32 -1.34 p$ & 0.91 & $2.14 + 0.81 p$ & & & 1 & 2.06 \\
 & & sB & $0.75 -0.74 p$ & $0.30 -1.40 p$ & 0.86 & $2.12 + 0.93 p$ & $1.49 -0.27 p$ & 3.7 &0.20 & 0.42\\
 & & sH & $0.85 -0.81 p$ & $0.27 -1.34 p$ & 0.82 & $2.32 + 0.81 p$ & & & 0.29 & 0.62 \\
 & $0.6 \theta_0$ & PL & $0.67 -0.82 p$ & $0.44 -1.44 p$ & 1.45 & $2.16 + 0.62 p$ & & & 1 & 1.90 \\
 & & sB & $0.74 -0.81 p$ & $0.23 -1.43 p$ & 1.52 & $2.31 + 0.62 p$ & $1.22 -0.12 p$ & 3.2 &  0.46 & 0.89 \\
 & & sH & $0.75 -0.82 p$ & $0.32 -1.42 p$ & 1.36 & $2.38 + 0.59 p$ & & & 0.44 & 0.84 \\
 & $0.8 \theta_0$ & PL & $0.59 -0.79 p$ & $0.43 -1.49 p$ & 2.19 & $1.99 + 0.54 p$ & & &  1 & 0.59 \\
 & &  sB & $0.59 -0.78 p$ & $0.42 -1.54 p$ & 2.29 & $2.08 + 0.52 p$ & $3.20 -0.59 p$ & 1.9 & 0.28 & 0.17\\
 & & sH & $0.66 -0.80 p$ & $0.28 -1.48 p$ & 2.19 & $2.18 + 0.50 p$ & & & 0.18 & 0.10 \\
 & $\theta_0$ & PL & $0.71 -0.79 p$ & $0.51 -1.56 p$ & 2.67 & $2.06 + 0.46 p$ & & & 1 & 1.30 \\
 & & sB & $0.71 -0.77 p$ & $0.46 -1.66 p$ & 3.07 & $2.09 + 0.45 p$ & $1.90 -0.35 p$ & 2.4 & 0.27 & 0.36 \\
 & & sH & $0.76 -0.80 p$ & $0.34 -1.55 p$ & 2.74 & $2.16 + 0.44 p$ & & & 0.24 & 0.33\\
\hline
 0.2 & 0 &  PL & $0.79 -0.80 p$ & $0.51 -1.44 p$ & 3.73 & $2.61 + 0.40 p$ & & & 1 & 0.16 \\
 & & sB & $0.78 -0.79 p$ & $0.45 -1.44 p$ & 3.77 & $2.58 + 0.42 p$ & $5.89 -0.96 p$ &  1.4 & 0.36 & 0.06 \\
 & & sH & $0.88 -0.80 p$ & $0.25 -1.39 p$ & 3.64 & $2.68 + 0.41 p$ & & & 1.57 & 0.23 \\
 & $0.2 \theta_0$ & PL & $0.79 -0.81 p$ & $0.53 -1.40 p$ & 3.75 & $2.63 + 0.37 p$ & & & 1 & 0.59 \\
 & & sB & $0.78 -0.78 p$ & $0.44 -1.48 p$ & 4.25 & $2.62 + 0.38 p$ & $2.69 -0.50 p$ & 2.1 &  0.05 & 0.03 \\
 & & sH & $0.88 -0.82 p$ & $0.23 -1.36 p$ & 3.87 & $2.72 + 0.37 p$ & & & 0.03 & 0.02 \\
 & $0.4 \theta_0$ & PL & $0.77 -0.85 p$ & $0.57 -1.36 p$ & 4.54 & $2.72 + 0.24 p$ & & & 1 & 1.71 \\
 & & sB & $0.83 -0.78 p$ & $-0.43 -2.22 p$ & 21.03 & $3.22 - 0.14 p$ & $0.34 -0.05 p$ & 8.3 & 0.06 & 0.11 \\
 & & sH & $0.88 -0.86 p$ & $0.32 -1.36 p$ & 4.86 & $2.90 + 0.19 p$ & & & 0.25 & 0.44 \\
 & $0.6 \theta_0$ & PL & $0.70 -0.86 p$ & $0.80 -1.45 p$ & 7.38 & $2.90 + 0.02 p$ & & & 1 & 1.07 \\
 & & sB & $0.74 -0.85 p$ & $0.70 -2.55 p$ & 24.27 & $3.14 - 0.35 p$ & $0.60 -0.10 p$ & 4.4 & 0.28 & 0.31 \\
 & & sH & $0.76 -0.87 p$ & $0.18 -1.37 p$ & 8.67 & $2.86 - 0.01 p$ & & & 0.36 & 0.39 \\
  & $0.8 \theta_0$ & PL & $0.62 -0.83 p$ & $0.64 -1.26 p$ & 9.42 & $2.62 + 0.00 p$ & & & 1 & 0.26 \\
 & & sB & $0.60 -0.81 p$ & $0.15 -2.14 p$ & 24.47 & $2.88 - 0.33 p$ & $1.25 -0.26 p$ & 2.7 & 0.05 & 0.01 \\
 & & sH & $0.61 -0.81 p$ & $-1.38 -0.86 p$ & 14.05 & $2.35 + 0.03 p$ & & & 0.05 & 0.01 \\
 & $\theta_0$ & PL & $0.73 -0.81 p$ & $0.63 -1.03 p$ & 6.96 & $2.29 + 0.20 p$ & & & 1 & 0.25 \\
 & & sB & $0.71 -0.79 p$ & $0.21 -1.50 p$ & 24.46 & $2.79 - 0.27 p$ & $1.44 -0.29 p$ & 3.9 & 0.14 & 0.03 \\
 & & sH & $0.72 -0.80 p$ & $-0.80 -0.77 p$ & 12.83 & $2.11 + 0.15 p$ & & & 0.13 & 0.03 \\
\hline
\end{tabular}
\caption{fit results for $\nu_{obs} < \nu_c$. \emph{PL}, \emph{sB} and \emph{sH} refer to different fit functions described in the text and $\alpha_0$, $\alpha_1$, $\tau_b$, $\bar{C}$ and $\sigma$ to their parameters. $^\dagger$Please note that the scale of the \emph{reduced} $\chi^2$ result is arbitrary because it depends on the number (85 or 60) and spacing (logarithmic in time) of synthetic light curve datapoints (chosen purely to properly resolve the light curve evolution) as well as an arbitrary error of 10 \% on each datapoint, which have no physical meaning.}
\label{powerlaw_fit_optical_table}
\end{table*}

\begin{table*}
\centering
\begin{tabular}{|ll|l|llllll|ll|}
\hline
$\theta_0$ (rad) & $\theta_{obs}$ & fit & $\alpha_0$ & $\alpha_1$ & $< \tau_b >$ & $^{10}\log \bar{C}$ & $\sigma$ & $<\tau_{0.9} / \tau_b>$ & $\chi^2 / \chi^2_{PL}$ & $\chi^2$, red. \\
\hline
 0.05 & 0 & PL & $0.50 -0.72 p$ & $0.20 -1.23 p$ & 0.08 & $-0.03 + 0.20 p$ & & &  1 & 1.92 \\
 & & sB & $0.50 -0.71 p$ & $0.20 -1.23 p$ & 0.07 & $-0.04 + 0.20 p$ & $23.81 -4.90 p$ &  1.1 & 1.00 & 1.94 \\
 & & sH & $0.93 -0.76 p$ & $0.20 -1.23 p$ & 0.06 & $0.34 + 0.19 p$ & & & 1.09 & 2.10 \\
 & $0.2 \theta_0$ & PL & $0.50 -0.74 p$ & $0.21 -1.23 p$ & 0.08 & $0.03 + 0.14 p$ & & & 1 & 1.86 \\
 & & sB & $0.49 -0.71 p$ & $0.22 -1.24 p$ & 0.08 & $-0.01 + 0.18 p$ & $10.25 -2.27 p$ & 1.4 & 0.97 & 1.82 \\
 & & sH & $0.95 -0.80 p$ & $0.21 -1.24 p$ & 0.06 & $0.43 + 0.12 p$ & & & 0.98 & 1.83 \\
 & $0.4 \theta_0$ & PL & $0.47 -0.82 p$ & $0.26 -1.27 p$ & 0.12 & $0.24 - 0.12 p$ & & & 1 & 2.29 \\
 & & sB & $0.52 -0.79 p$ & $0.27 -1.27 p$ & 0.11 & $0.24 - 0.05 p$ & $5.26 -0.99 p$ & 1.7 & 0.92 & 2.14 \\
 & & sH & $0.89 -0.88 p$ & $0.27 -1.27 p$ & 0.09 & $0.72 - 0.14 p$ & & & 0.88 & 2.00 \\
 & $0.6 \theta_0$ & PL & $0.33 -0.80 p$ & $0.33 -1.31 p$ & 0.20 & $0.15 - 0.28 p$ & & & 1 & 2.17 \\
 & & sB & $0.33 -0.80 p$ & $0.33 -1.31 p$ & 0.19 & $0.15 - 0.26 p$ & $13.33 -2.63 p$ & 1.3 & 0.99 & 2.17 \\
 & & sH & $0.66 -0.85 p$ & $0.33 -1.32 p$ & 0.14 & $0.67 - 0.33 p$ & & & 1.06 & 2.31 \\
 & $0.8 \theta_0$ & PL & $0.32 -0.77 p$ & $0.39 -1.36 p$ & 0.27 & $0.01 - 0.33 p$ & & & 1 & 2.37 \\
 & & sB & $0.31 -0.75 p$ & $0.39 -1.36 p$ & 0.26 & $-0.02 - 0.30p$ & $11.95 -2.67 p$ & 1.3 & 0.98 & 2.34 \\
 & & sH & $0.60 -0.81 p$ & $0.39 -1.37 p$ & 0.21 & $0.54 - 0.41 p$ & & & 1.03 & 2.44 \\
 & $\theta_0$ & PL & $0.45 -0.78 p$ & $0.46 -1.42 p$ & 0.35 & $-0.06 - 0.40 p$ & & & 1 & 2.55 \\
 & & sB & $0.43 -0.76 p$ & $0.46 -1.42 p$ & 0.33 & $-0.08 - 0.36 p$ & $7.98 -1.67 p$ & 1.4 & 0.96 & 2.47 \\
 & & sH & $0.66 -0.81 p$ & $0.45 -1.42 p$ & 0.29 & $0.34 - 0.45 p$ & & & 0.97 & 2.45 \\
\hline
 0.1 & 0 & PL & $0.50 -0.75 p$ & $0.27 -1.29 p$ & 0.52 & $0.32 - 0.41 p$ & & & 1 & 0.60 \\
 & & sB & $0.50 -0.74 p$ & $0.28 -1.29 p$ & 0.51 & $0.31 - 0.40 p$ & $12.62 -2.52 p$ &  1.2 & 0.95 & 0.58 \\
 & & sH & $0.66 -0.77 p$ & $0.25 -1.28 p$ & 0.44 & $0.60 - 0.44 p$ & & & 1.40 & 0.84 \\
 & $0.2 \theta_0$ & PL & $0.49 -0.75 p$ & $0.27 -1.29 p$ & 0.55 & $0.30 - 0.43 p$ & & & 1 & 0.65 \\
 & & sB & $0.49 -0.74 p$ & $0.30 -1.31 p$ & 0.53 & $0.34 - 0.42 p$ & $5.96 -1.26 p$ & 1.6 & 0.67 & 0.44 \\
 & & sH & $0.65 -0.78 p$ & $0.26 -1.29 p$ & 0.48 & $0.64 - 0.49 p$ & & & 0.69 & 0.44 \\
 & $0.4 \theta_0$ & PL & $0.49 -0.80 p$ & $0.34 -1.32 p$ & 0.76 & $0.51 -0.66 p$ & & & 1 & 1.64 \\
 & & sB & $0.49 -0.74 p$ & $0.41 -1.39 p$ & 0.68 & $0.51 - 0.56 p$ & $2.53 -0.53 p$ & 3.0 & 0.38 & 0.63 \\
 & & sH & $0.66 -0.82 p$ & $0.33 -1.33 p$ & 0.65 & $0.88 -0.71 p$ & & & 0.33 & 0.55 \\
 & $0.6 \theta_0$ & PL & $0.43 -0.81 p$ & $0.48 -1.42 p$ & 1.26 & $0.63 -0.91 p$ & & & 1 & 1.50 \\
 & & sB & $0.48 -0.81 p$ & $0.44 -1.44 p$ & 1.25 & $0.70 -0.90 p$ & $2.22 -0.32 p$ &  2.6 & 0.63 & 0.96 \\
 & & sH & $0.56 -0.84 p$ & $0.43 -1.42 p$ & 1.14 & $0.93 -0.96 p$ & & & 0.50 & 0.76 \\
 & $0.8 \theta_0$ & PL & $0.35 -0.79 p$ & $0.54 -1.49 p$ & 2.00 & $0.38 -0.99 p$ & & & 1 & 0.40 \\
 & & sB & $0.34 -0.78 p$ & $0.59 -1.53 p$ & 2.04 & $0.40 -0.99 p$ & $6.73 -1.53 p$ & 1.6 & 0.56 & 0.22 \\
 & & sH & $0.44 -0.81 p$ & $0.42 -1.48 p$ & 1.96 & $0.57 -1.02 p$ & & & 0.81 & 0.30 \\
 & $\theta_0$ & PL & $0.48 -0.79 p$ & $0.71 -1.58 p$ & 2.52 & $0.44 - 1.08 p$ & & & 1 & 0.82 \\
 & & sB & $0.46 -0.77 p$ & $0.74 -1.66 p$ & 2.69 & $0.37 - 1.05 p$ & $3.48 -0.76 p$ & 2.1 & 0.39 & 0.32 \\
 & & sH & $0.54 -0.80 p$ & $0.49 -1.54 p$ & 2.50 & $0.49 -1.07 p$ & & & 0.28 & 0.23 \\
\hline
 0.2 & 0 & PL & $0.56 -0.80 p$ & $0.52 -1.45 p$ & 3.66 & $0.83 -1.11 p$ & & & 1 & 0.14 \\
 & & sB & $0.55 -0.79 p$ & $0.49 -1.44 p$ & 3.65 & $0.80 -1.09 p$ & $9.44 -1.85 p$ & 1.3 & 0.61 & 0.09 \\
 & & sH & $0.67 -0.80 p$ & $0.28 -1.40 p$ & 3.48 & $0.95 -1.10 p$ & & & 2.72 & 0.36 \\
 & $0.2 \theta_0$ & PL & $0.57 -0.81 p$ & $0.53 -1.41 p$ & 3.65 & $0.87 - 1.14 p$ & & & 1 & 0.40 \\
 & & sB & $0.54 -0.78 p$ & $0.54 -1.48 p$ & 3.96 & $0.84 -1.13 p$ & $4.07 -0.85 p$ &  1.9 & 0.10 & 0.04 \\
 & & sH & $0.67 -0.82 p$ & $0.26 -1.37 p$ & 3.67 & $0.98 -1.14 p$ & & & 0.15 & 0.05 \\
 & $0.4 \theta_0$ & PL & $0.57 -0.85 p$ & $0.66 -1.39 p$ & 4.28 & $1.11 - 1.33 p$ & & & 1 & 1.28 \\
 & & sB & $0.58 -0.78 p$ & $1.21 -2.55 p$ & 18.24 & $1.47 - 1.70 p$ & $0.55 -0.11 p$ & 8.6 & 0.06 & 0.08 \\
 & & sH & $0.68 -0.86 p$ & $0.33 -1.35 p$ & 4.55 & $1.18 - 1.33 p$ & & & 0.19 & 0.26 \\
 & $0.6 \theta_0$ & PL & $0.46 -0.86 p$ & $0.62 -1.36 p$ & 7.33 & $0.96 -1.48 p$ & & & 1 & 0.72 \\
 & & sB & $0.51 -0.85 p$ & $1.03 -2.48 p$ & 24.07 & $1.10 - 1.83 p$ & $0.74 -0.13 p$ & 4.6 & 0.29 & 0.21 \\
 & & sH & $0.54 -0.87 p$ & $0.12 -1.37 p$ & 9.25 & $0.96 -1.52 p$ & & & 0.33 & 0.24 \\
 & $0.8 \theta_0$ & PL & $0.40 -0.82 p$ & $0.45 -1.21 p$ & 9.71 & $0.64 - 1.48 p$ & & & 1 & 0.15 \\
 & & sB & $0.39 -0.81 p$ & $0.42 -2.14 p$ & 24.29 & $0.92 -1.84 p$ & $1.63 -0.36 p$ & 2.5 & 0.09 & 0.01 \\
 & & sH & $0.39 -0.81 p$ & $-2.03 -0.72 p$ & 16.11 & $0.31 -1.47 p$ & & &  0.09 & 0.01 \\ 
 & $\theta_0$ & PL & $0.50 -0.81 p$ & $0.42 -1.01 p$ & 7.82 & $0.32 -1.30 p$ & & & 1 & 0.13 \\
 & & sB & $0.48 -0.80 p$ & $0.12 -1.48 p$ & 24.44 & $0.83 -1.79 p$ & $2.03 -0.45 p$ & 3.3 & 0.08 & 0.01 \\
 & & sH & $0.49 -0.80 p$ & $-1.70 -0.58 p$ & 16.48 & $0.07 -1.37 p$ & & & 0.06 & 0.01 \\
\hline
\end{tabular}
\caption{Same as table \ref{powerlaw_fit_optical_table}, now for $\nu_{obs} > \nu_c$.}
\label{powerlaw_X-rays_table}
\end{table*}

Tables \ref{powerlaw_fit_optical_table} and \ref{powerlaw_X-rays_table} show the results of the analysis of light curves with different $p$, $\theta_{obs}$ and $\theta_0$ values using different power law descriptions. Each data point of the light curves, consisting of 85 data points per curve between observer times $10^{-2}$ days and 26 days for $\theta_0 = 0.05$ rad. and $\theta_0 = 0.1$ rad. and 60 per curve between observer times $10^{-1}$ days and 26 days for $\theta_0 = 0.2$ rad., was given an error of ten percent and three different jet break functions were fitted using a least squares algorithm. A baseline frequency $\nu = 4.56 \times 10^{14}$ Hz (R-band) was used for table \ref{powerlaw_fit_optical_table} and a baseline frequency $\nu = 5 \times 10^{17}$ Hz (2.07 KeV) for table \ref{powerlaw_X-rays_table}. The fit function for a sharp power law, labeled ``PL'' in the table and below, is given by 
\begin{equation}
 \bar{F}(\tau) = \left\{ \begin{array}{cl} \bar{C} (\tau / \tau_b)^{\alpha_0}, & \tau < \tau_b, \\ 
  \bar{C} (\tau / \tau_b)^{\alpha_1}, & \tau > \tau_b \end{array} \right. .
\end{equation}
The fit function for a smooth power law transition, equivalent to that used by \citealt{Beuermann1999} and labeled ``sB'', is given by
\begin{equation}
 \bar{F}(\tau) = \bar{C} \left[ \left( \frac{\tau}{\tau_b} \right)^{- \alpha_0 \sigma} + \left( \frac{\tau}{\tau_b} \right)^{-\alpha_1 \sigma} \right]^{-1/\sigma}.
\label{Beuermann_fit_function_equation}
\end{equation}
The alternative smooth power law transition fit function, labeled ``sH'' is the same as the one used by \cite{Harrison1999} and given by
\begin{equation}
 \bar{F}(\tau) = \bar{C} \left\{ 1 - \exp[ -(\tau / \tau_b)^{\alpha_0 - \alpha_1}] \right\} (\tau / \tau_b)^{\alpha_1}.
\end{equation}
In this equation the pre-break power law slope is retrieved from the Taylor series of the exponential term.

The different fit variable results are represented in the tables as follows. Since the fit results confirm that the slopes $\alpha_0$ and $\alpha_1$ linearly depend on $p$ (as shown in Figs. \ref{alpha_plot} and \ref{alphaX_plot} for fits using sharp power laws), the entries contain this linear dependence as determined from the full range of $p$ fits rather than a values for each individual $p$. The logarithm of the numerical scale factor $\bar{C}$ and the sharpness of the smooth power law fit also depend linearly on $p$, and are presented in the same fashion. The break time $\tau$ depends only weakly on $p$ and is represented by its average value $<\tau>$, weighing equally all individual $p$ value fits. We also give the reduced $\chi^2$ value of each fit, again averaged over the different $p$ value fits, as well as the ratio of the unreduced $\chi^2$ of each fit to that of a sharp power law fit. For the latter, these ratios were calculated before the average was taken. We emphasize that by themselves, the reduced $\chi^2$ results are to some degree arbitrary, since they depend on arbitrary quantities like the number of datapoints in a synthetic light curve, the spacing of these datapoints and an artificial ten percent error on each datapoint, and that they should be interpreted only in a relative sense.

Using the prescriptions from table \ref{characteristics_table} and equations \ref{flux_equations}, the flux for any combination of parameter values can be reproduced from the fit results in tables \ref{powerlaw_fit_optical_table} and \ref{powerlaw_X-rays_table}. For $\nu < \nu_c$, we get:
\begin{eqnarray}
 F_G & = & \frac{(1+z)^{(3-p)/2}}{d_{28}^2} \frac{p-1}{3p-1} \left( \frac{p-2}{p-1} \right)^{p-1} \epsilon_e^{p-1} \epsilon_B^{(p+1)/4} \xi_N^{2-p} \times \nonumber \\ 
  & & \kappa^1 \lambda^{(p+1)/4} \left( \frac{\nu_{\oplus}}{4.56 \times 10^{14} \textrm{ Hz}} \right)^{(1-p)/2} \bar{F} (\tau) \textrm{ mJy},
\end{eqnarray}
for the ISM case and $\bar{F}$ referring to fit results from table \ref{powerlaw_fit_optical_table}. We have now added a `$\oplus$' to the frequency to emphasize that this frequency is expressed in the observer frame, like the characteristic frequencies in table \ref{characteristics_table} and the frequencies in Eq. \ref{flux_equations}, and related to the frequency $\nu$ in the burster frame via $\nu_{\oplus} = \nu / (1+z)$. Note that $\tau$ is still expressed in the burster frame, in order to keep the characteristic functions (and hence the power law fit results) redshift-independent. For $\nu > \nu_c$ we have:
\begin{eqnarray}
 F_H & = & \frac{(1+z)^{(2-p)/2}}{d_{28}^2} \frac{p-1}{3p-1} \left( \frac{p-2}{p-1} \right)^{p-1} \epsilon_e^{p-1} \epsilon_B^{(p-2)/4} \xi_N^{2-p} \times \nonumber \\ 
 & & \kappa^{2/3} \lambda^{(3p-2)/12} \left( \frac{\nu_{\oplus}}{5 \times 10^{17} \textrm{ Hz}} \right)^{-p/2} \bar{F} (\tau) \textrm{ mJy},
\end{eqnarray}
for the ISM case and $\bar{F}$ referring to fit results from table \ref{powerlaw_X-rays_table}.

\subsection{Implications for the light curve slope}

\begin{figure}
 \centering
  \includegraphics[width=\columnwidth]{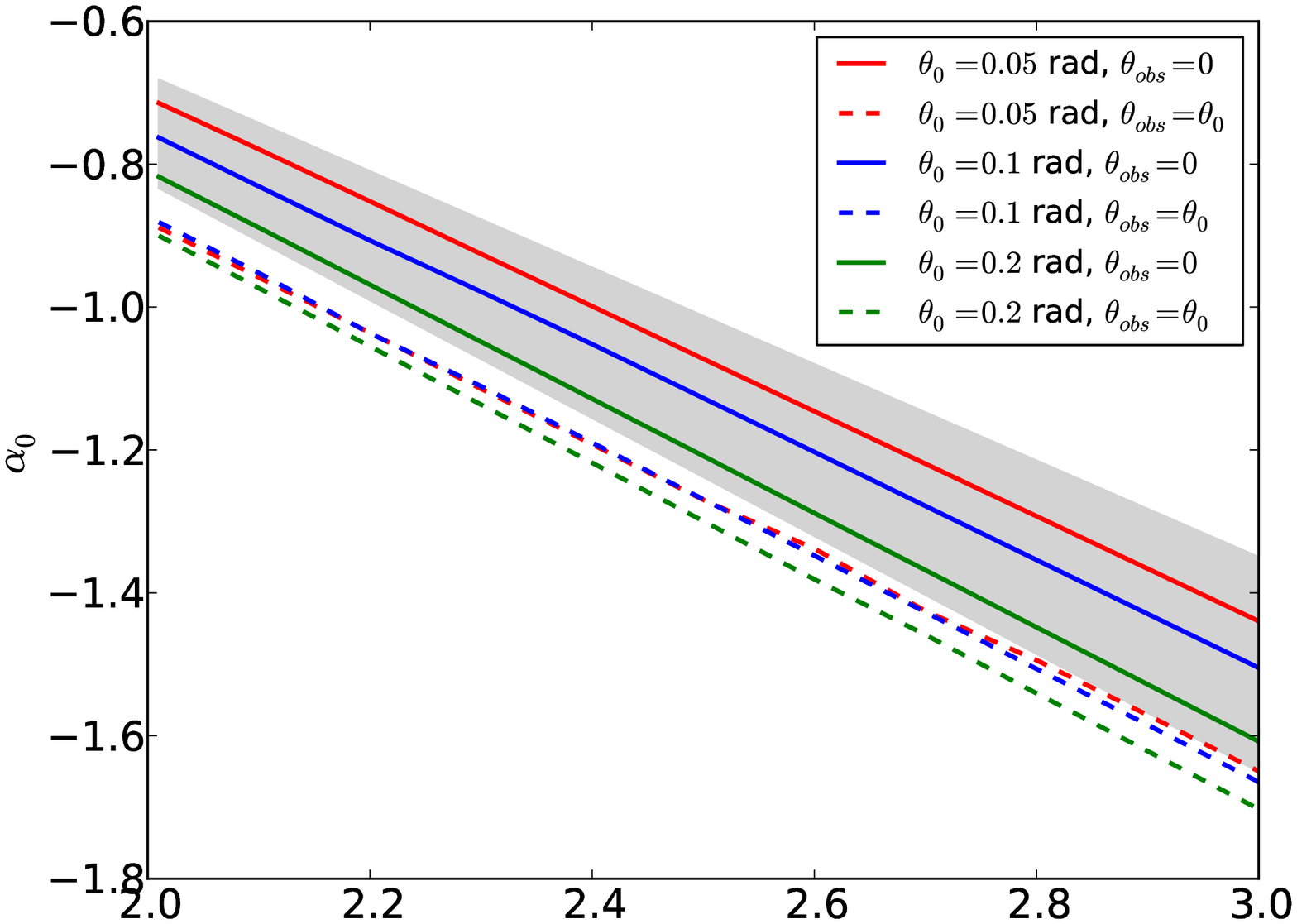}
  \includegraphics[width=\columnwidth]{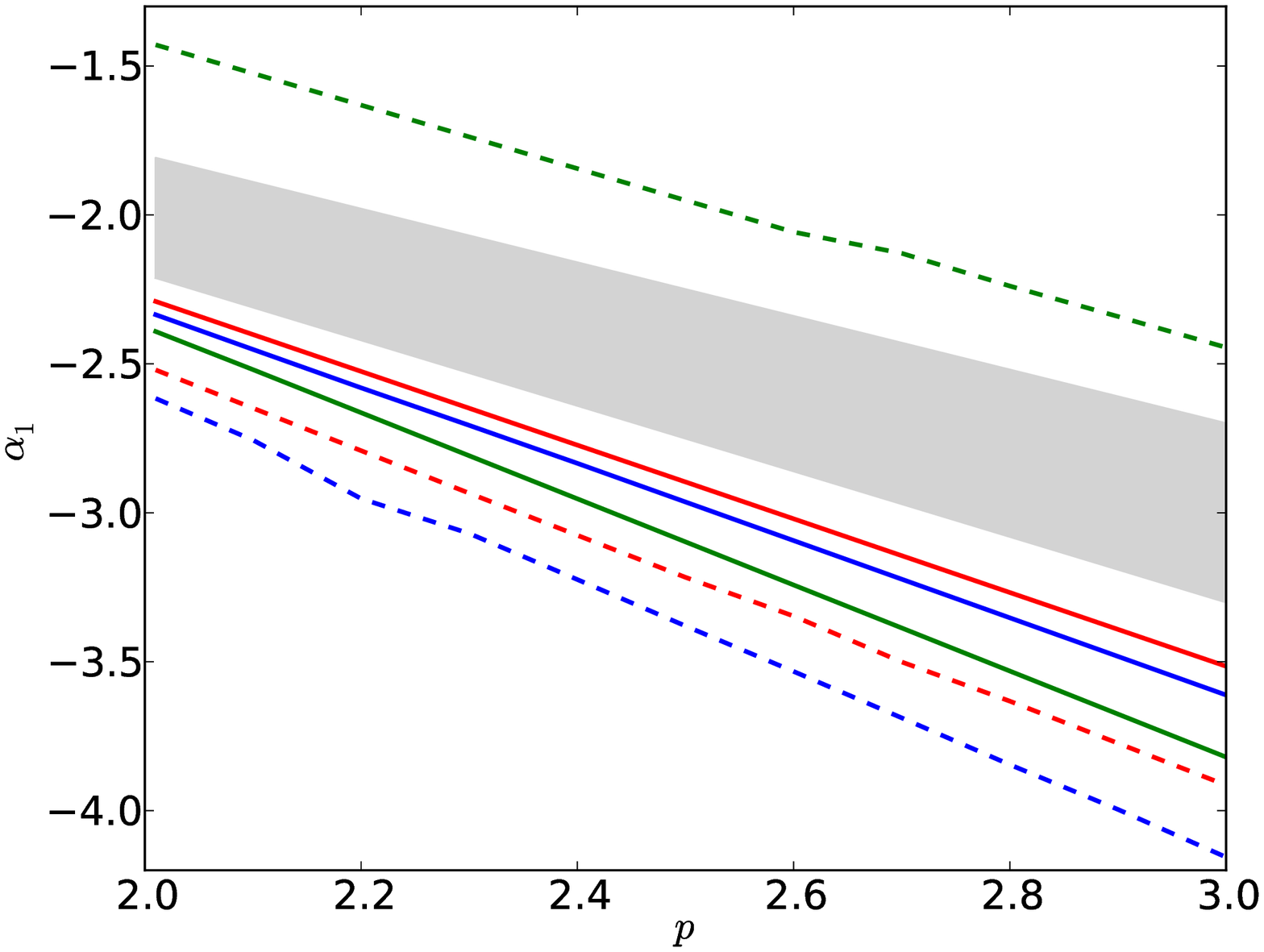}
  \caption{Pre-break temporal index $\alpha_0$ (top plot) and post-break temporal index $\alpha_1$ (bottom plot) for $\nu < \nu_c$ and for on-axis and on-edge observations of the three jet angles $\theta_0 = 0.05$, $0.1$, $0.2$ rad., according to sharp power law fits to synthetic light curves. The grey bands indicate the region within ten percent of the theoretically expected values, $3(1-p)/4$ and $-p$ for pre- and post-break respectively.}
\label{alpha_plot}
\end{figure}

\begin{figure}
 \centering
  \includegraphics[width=\columnwidth]{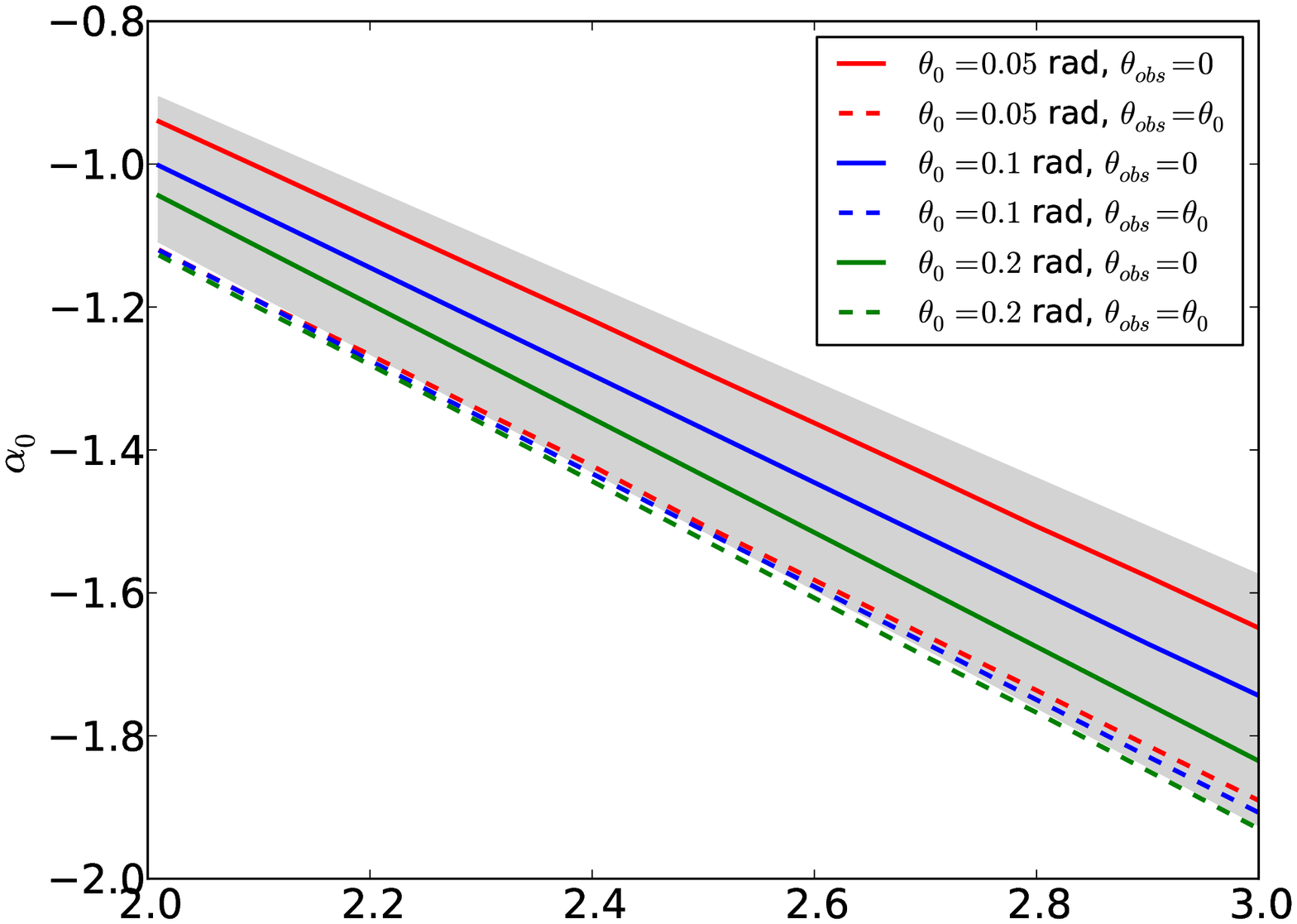}
  \includegraphics[width=\columnwidth]{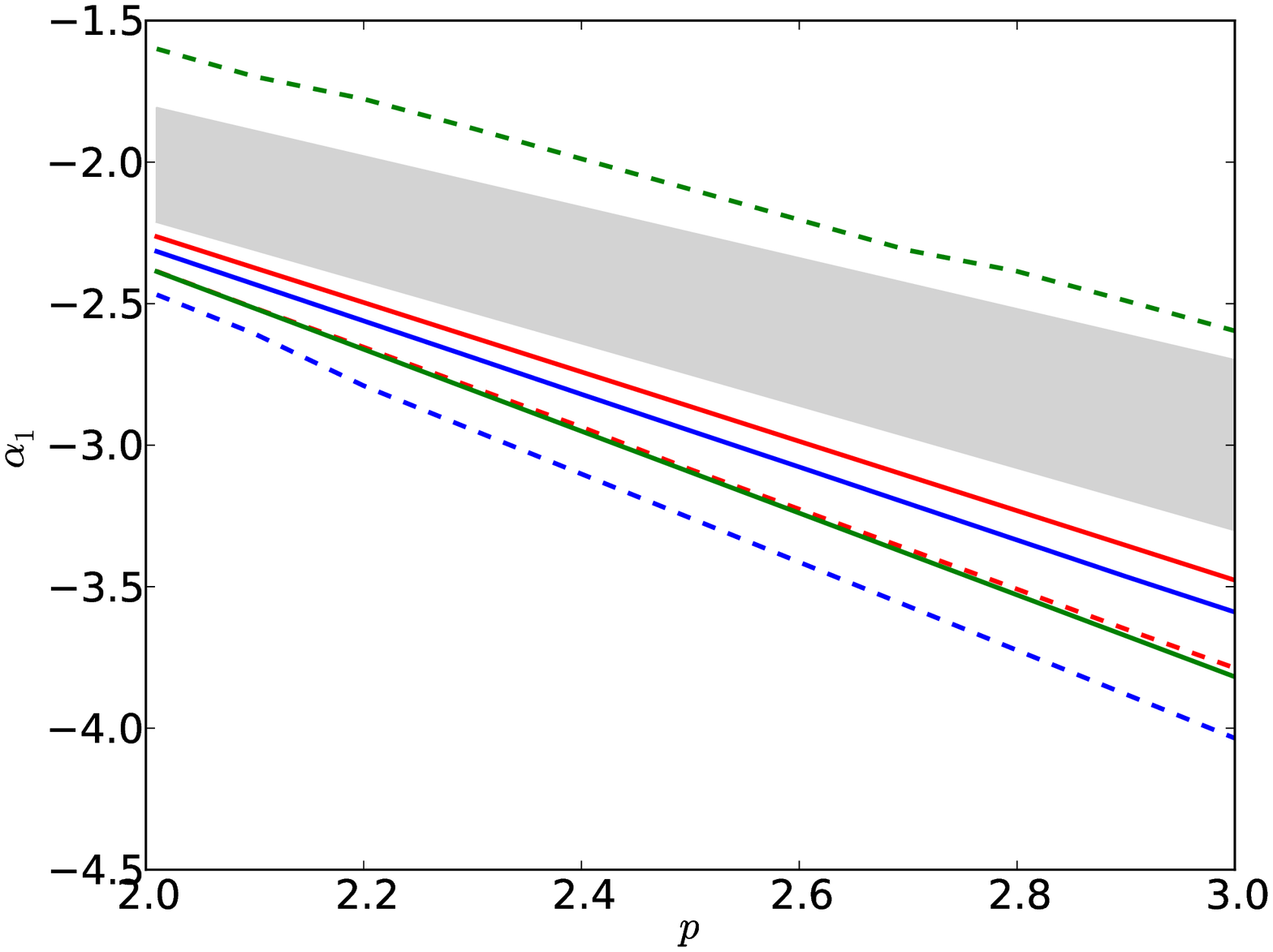}
  \caption{Same as Fig. \ref{alpha_plot}, now for $\nu > \nu_c$.}
\label{alphaX_plot}
\end{figure}

The predicted on-axis pre-break slopes are $3(1-p)/4$ for $\nu < \nu_c$ and $(2-3p)/4$ for $\nu > \nu_c$ and the tables show that these values are well reproduced by straight power law fits for $\theta_0 = 0.05$ rad. and $\theta_0 = 0.1$ rad. and reasonably well for $\theta_0 = 0.2$ rad. This can also be seen from the top plots in Figs. \ref{alpha_plot} and \ref{alphaX_plot}, that show $\alpha_0$ for each $p$ value and on-axis and on-edge observers.
  
\begin{figure}
 \centering
  \includegraphics[width=\columnwidth]{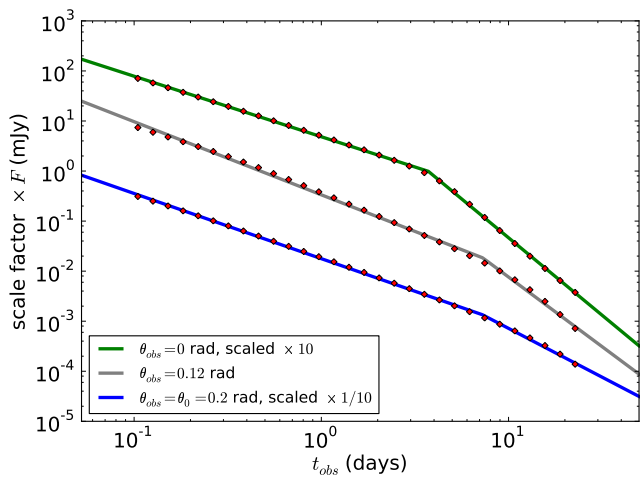}
  \caption{A comparison of sharp power law fits and synthetic light curves at observer angles $\theta_{obs} = 0$, $0.12$, $0.2$, rad. (top to bottom) for $\theta_0 = 0.2$ rad. Plotted is the case $\nu < \nu_c$. Other parameters are set as follows: $p = 2.5$, $\epsilon_e = 0.1$, $\epsilon_B = 0.01$, $\xi_N = 1.0$, $z = 0$, $d_{28} = 1$, $n_0 = 1$ cm$^{-3}$, $E_{iso} = 10^{53}$ erg. For clarity of presentation, only half the data points of the synthetic light curves are plotted. Two of the three curves have been scaled by a factor ten, again for presentation purposes.}
\label{observer_angle_plot}
\end{figure}

The post-break slopes, on the other hand, are \emph{not} consistent with the theoretically expected temporal index $-p$ \citep{Sari1999} nor with very smooth gradual transitions \citep{Kumar2000, Wei2000, Wei2002}, as can be seen from the tables and the bottom plots in Figs. \ref{alpha_plot} and \ref{alphaX_plot} (see also section \ref{transition_duration_subsection}) and are steeper to the extent that they fall well outside even a ten percent margin of the theoretical value. We find a steepening of about $-(0.5 + 0.5p)$ below the cooling break and about $-(0.25 + 0.5p)$ above the cooling break, both leading to post-break slopes of roughly $0.25 - 1.3 p$, although different observer angles, jet opening angles and heuristic descriptions of the break introduce a wide range of temporal indices. This confirms earlier numerical work, and was first shown from high-resolution simulations and for on-axis observers by \cite{Zhang2009} and for off-axis observers by \cite{vanEerten2010offaxis}. However, due to the vast increase in numerical resolution provided by the boosted frame approach, this is the first time the post-break slopes have been determined from simulations where the jet break is fully resolved, and the current values can be considered quantitatively accurate. These slopes should be compared to observational data, such as the systematic study of \emph{Swift} X-ray afterglows presented by \cite{Racusin2009}, that show a post-break slope for their sample of afterglows exhibiting `prominent jet breaks' that centers around $\alpha_1 \sim 2$. This difference in slopes means that it is exceedingly difficult, at least for the \emph{Swift} sample and at least for on-axis observers, to reconcile the data with a model of an initially top-hat blast wave decelerating into a constant medium. Even for off-axis observers this is becoming problematic, although a number of caveats apply: the jet break might be simply post-poned beyond what \emph{Swift} can observe \citep{vanEerten2010offaxis, vanEerten2011hiddenswift}, or only a fraction of an off-axis jet break is seen (see also Fig. \ref{observer_angle_plot}, discussed below). Instead, the post-break slopes are more consistent with the values normally associated with high-latitude emission (`region I' of the `canonical light curve', see \citealt{BingZhang2006, Racusin2009}), without necessarily implying that these should be interpreted as such, since this interpretation would require extremely narrow jets (embedded in quasi-spherical outflow in order to get regions II-IV of the canonical light curve) and simulations of jets with $\theta_0 \ll 0.05$ rad., the smallest angle discussed in this paper.
Possibly, \emph{Swift} GRBs do not predominantly explode into a homogeneous medium but in a different environment (e.g. stellar wind instead). Alternatively, the jet break might be hidden from view by an additional physical process, such as prolonged injection of energy into the blast wave (see e.g. \citealt{Nousek2006, Panaitescu2006, ZhangBing2006, Panaitescu2012}).

In Fig. \ref{observer_angle_plot} we show light curves and sharp power law fit results for $\theta_{obs} = 0.2$ rad. These illustrate the extent to which sharp power law fits overlap with the data. In practice, it is not very difficult for off-axis observations to push the final turnover associated with the jet break out in time beyond the timespan typically covered by \emph{Swift} (i.e. 10 days), especially once nonzero values for redshift $z$ are considered, which can lead either to a \emph{missing jet break} or a steepening that is far more shallow if detected at all because only the early part of the jet break is covered. An example of a jet break that is not fully detected is shown by the green dashed curve in Fig. \ref{alphaX_plot}, for $\theta_{obs} = \theta_0 = 0.2$ rad. However, in order to properly quantify these effects for e.g. \emph{Swift}, an approach is required that includes not just synthetic light curves but also accurately models instrument biases and expected measurement errors, similar to the one taken by \cite{vanEerten2010offaxis, vanEerten2011hiddenswift}. This will be the topic of a future study.

\begin{figure}
 \centering
  \includegraphics[width=\columnwidth]{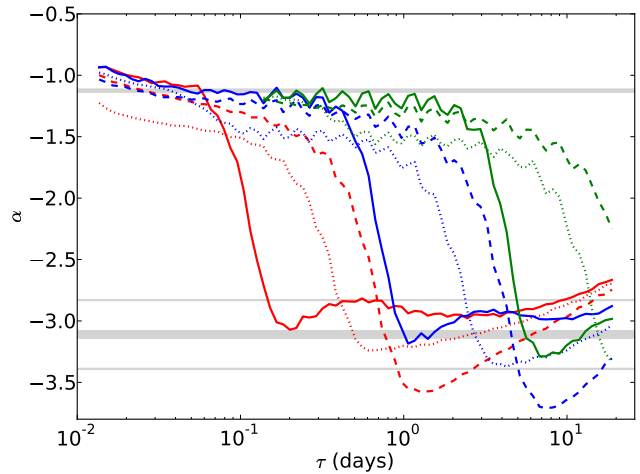}
  \includegraphics[width=\columnwidth]{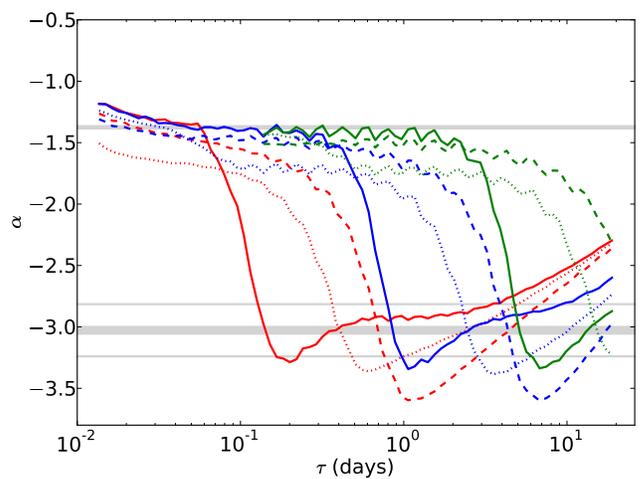}
  \caption{Temporal index evolution for $\nu < \nu_c$ (top plot) and $\nu > \nu_c$ (bottom plot), for $p = 2.5$. $\theta_0 = 0.05$, $0.1$, $0.2$ rad and $\theta_{obs} = 0$, $0.6 \theta_0$, $\theta_0$, using the same colors and line styles as in Fig. \ref{characteristics_plot}. The top grey lines indicate the theoretical pre-break value. The narrow bottom grey lines indicate the ranges of sharp power law values found for the post-break slope, for all opening angles and observer angles except $\theta_{obs} > 0.4 \theta_0$ with $\theta_0 = 0.2$ rad, where the temporal index did not reach a minimum during before 26 days. The thick bottom grey lines denote the middle of these ranges.}
\label{slopes_plot}
\end{figure}

\begin{figure}
 \centering
  \includegraphics[width=\columnwidth]{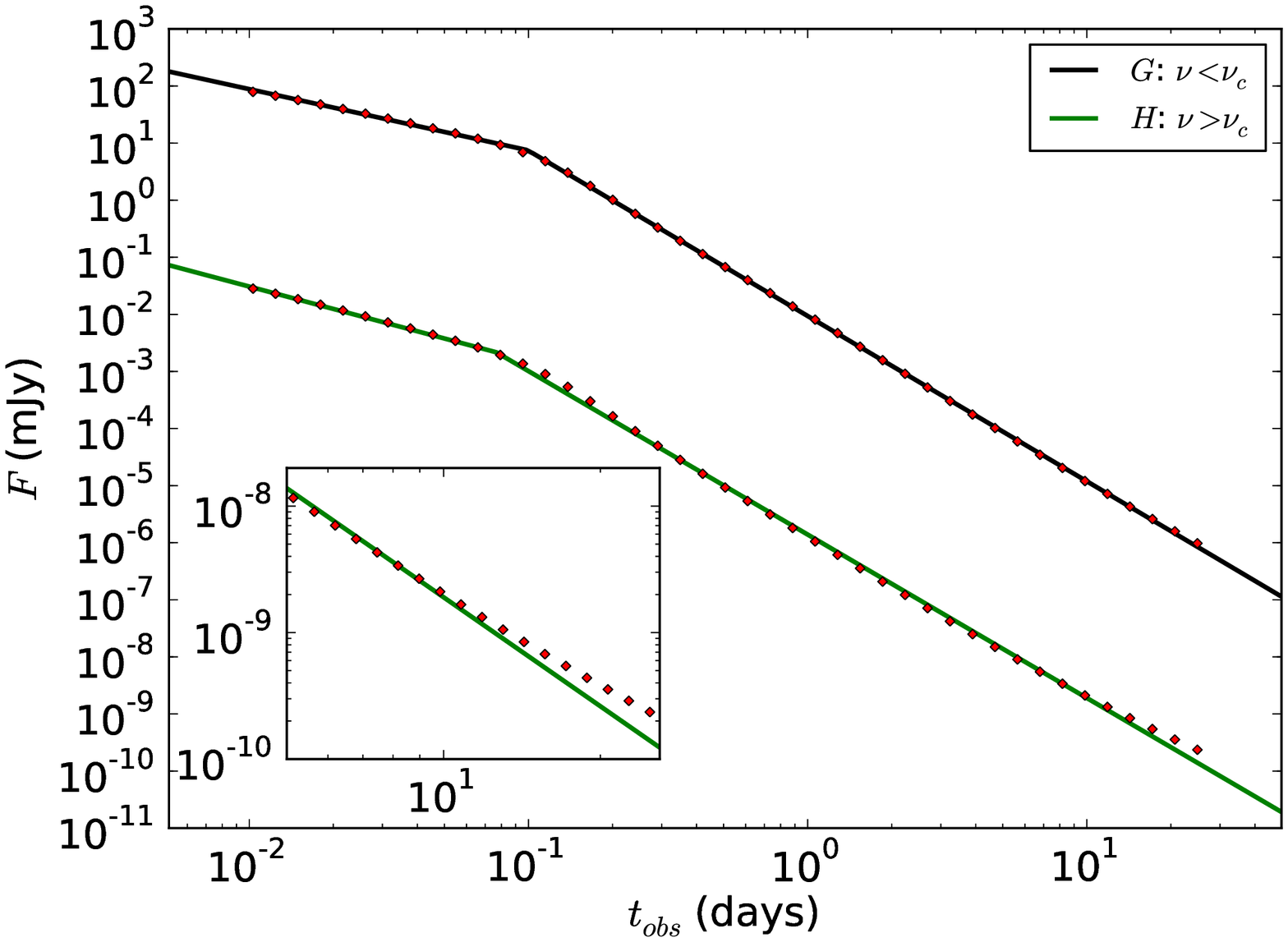}
  \caption{A comparison between light curves for $\nu > \nu_c$ and $\nu < \nu_c$ and $\theta_0 = 0.05$ rad, $\theta_{obs}$ = 0.0. Other parameters are set as follows: $p = 2.5$, $\epsilon_e = 0.1$, $\epsilon_B = 0.01$, $\xi_N = 1.0$, $z = 0$, $d_28 = 1$, $n_0 = 1$ cm$^{-3}$, $E_{iso} = 10^{53}$ erg. For clarity of presentation, only half the data points of the synthetic light curves are plotted. The inset plot shows a zoom in of the late time X-ray curve, without skipping data points of the synthetic light curve.}
\label{XvsO_plot}
\end{figure}

After the onset of the jet break, the time evolution of the light curves in general does not follow a single power law evolution, as can be seen from Fig. \ref{slopes_plot}. Given that the synthetic light curves consist of 85 (60) datapoints and were given artificial measurement errors of ten percent, the reduced $\chi^2$ values for the various fits reported in tables \ref{powerlaw_fit_optical_table} and \ref{powerlaw_X-rays_table} demonstrate that power law fit functions nevertheless fit the light curve surprisingly well. Even the $\nu > \nu_c$ fits for $\theta_0 = 0.05$ rad. have a small reduced $\chi^2$ value. The reason that these are nevertheless noticeably higher than the other fits can be inferred from the late time behavior of the temporal indices for the narrow jet in Fig. \ref{slopes_plot}. Above the cooling break, the emission is dominated by a smaller region closer to the shock front than is the case below the cooling break. As a result, the observed flux above the cooling break at any given time consists of contributions from a smaller timespan in emission times. It will therefore take less time for a change in the nature of the evolution of the blast wave to become noticeable than for flux below the cooling break, as illustrated by the comparison shown in Fig. \ref{XvsO_plot}. What is seen for the $\nu > \nu_c$ curve at late times is the onset of the transition to the non-relativistic regime, a consequence of the fact that the smaller the opening angle, the smaller the total energy in the jets (with energy  in jet and counterjet $E_j \approx E_{iso} \theta_0^2 / 2$).

\subsection{Implications for the break times}

\begin{figure}
 \centering
  \includegraphics[width=\columnwidth]{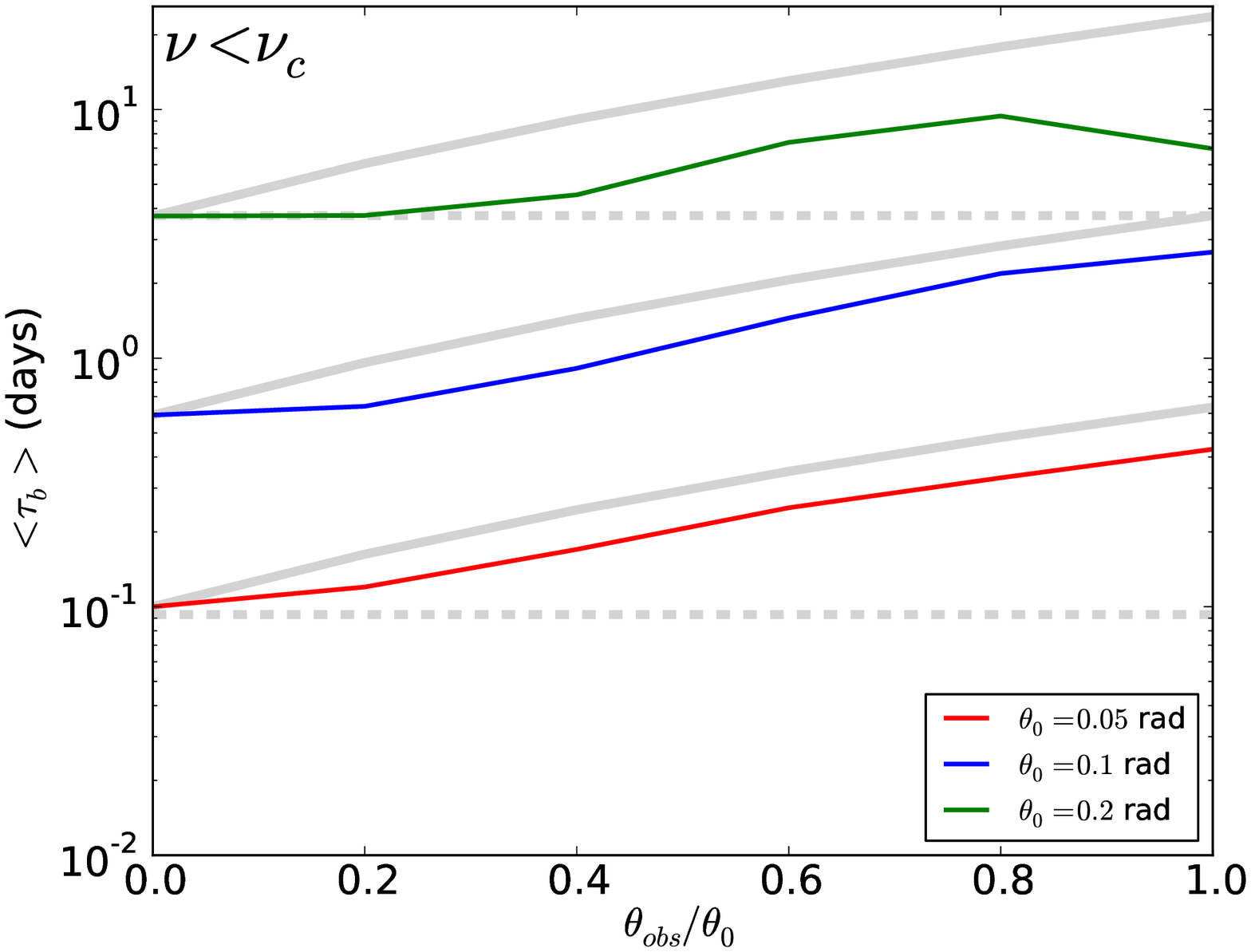}
  \includegraphics[width=\columnwidth]{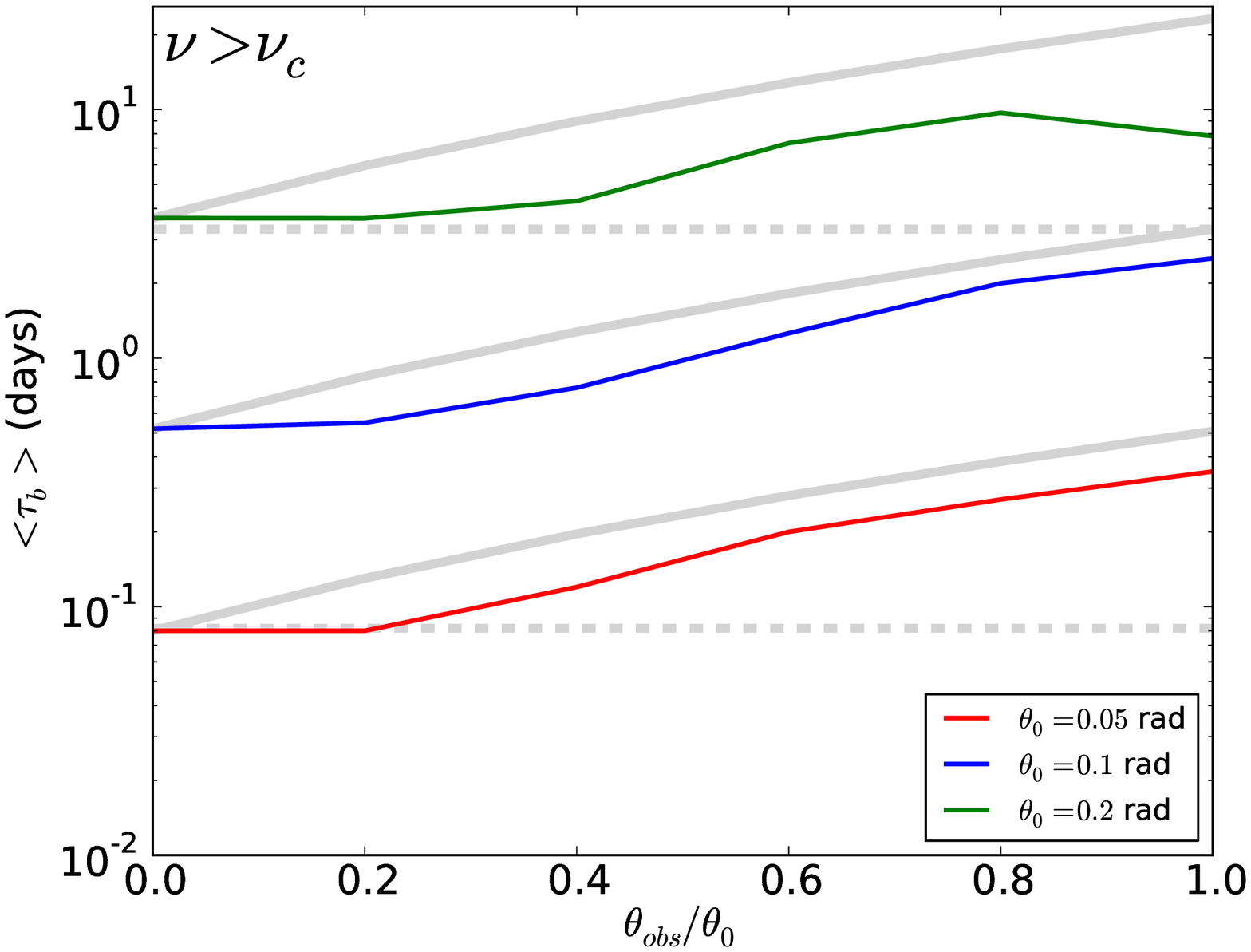}
  \caption{Jet break times averaged over a range of $p$ values for $\nu < \nu_c$ (top plot) and $\nu > \nu_c$, as determined from sharp power law fits for different observer angles $\theta_{obs}$ and different jet opening angles $\theta_0$. The solid grey curves indicate $\tau \propto (\theta_0 + \theta_{obs})^{8/3}$. The dashed grey lines indicate break times for an on-axis observer, scaled from the on-axis break time  $\tau_{0.05}$ for $\theta_0 = 0.05$ rad., using $\tau = \tau_{0.05} (\theta_{0} / 0.05)^{8/3}$.}
\label{break_times_plot}
\end{figure}

The evolution of the jet break time, as determined using a sharp power law fit, is shown in Fig. \ref{break_times_plot}. If there were no lateral spreading at all, the jet break would be determined completely by the different edges becoming visible, and as a result the onset $\tau_{b0}$ and end $\tau_{b1}$ of the jet break would be given by $\tau_{b0} \propto (\theta_0 - \theta_{obs})^{8/3}$ and $\tau_{b1} \propto (\theta_0 + \theta_{obs})^{8/3}$ respectively \citep{vanEerten2010offaxis}. For a jet observed on-edge, the nearest edge is visible already at $\tau = 0$, while the relative angle of the far edge is at its maximum distance of $2 \theta_0$. In reality, jet break is influenced by jet spreading as well. Also, the intermediate light curve slope change at the onset of the break is not as steep as the final slope change at the end of the break even for pure radial flow. These facts, together with the fact that the onset of the break is usually sufficiently early to be overwhelmed in light curve data (e.g. from \emph{Swift}) by other early time features such as flares or plateaus, render it likely that in practice it is the end of the jet break rather than the onset of the jet break that will be captured by a broken power law fit to the data. The relation between measured break time and jet opening angle will therefore lie closer to $\tau_{b} \propto (\theta_0 + \theta_{obs})^{8/3}$, than to  $\tau_{b} \propto (\theta_0)^{8/3}$, for general observer angle. Although the inferred jet breaks for the synthetic light curves do not fully reach this upper limit, Fig. \ref{break_times_plot} shows that, when the observer moves noticeably off-axis, they do trace this expected behavior at least for $\theta_0 = 0.05$ rad. and $\theta_0 = 0.1$ rad. The jet break time as a function of observer angle is very noisy for the wide jet with $\theta_0 = 0.2$ rad., mainly because in this case the jet break for observers far off-axis is not fully covered within the timespan of 26 days. For small observer angles, when both onset and end of the break are still fairly close to each other, the two breaks have not yet fully separated and the turnover is still described by a single smooth break centered at $\tau \propto (\theta_{obs})^{8/3}$, as indicated by Fig. \ref{break_times_plot} and the drop in $\sigma$ values for \emph{sB} fits between for increasing $\theta_{obs}$ (as shown in tables \ref{powerlaw_fit_optical_table} and \ref{powerlaw_X-rays_table}).

\subsection{Implications for the transition duration}
\label{transition_duration_subsection}

The parameter $\sigma$ in \emph{sB} type fits is a measure of the sharpness of the transition. From $\sigma$ a measure for the duration of the jet break transition can be derived as follows. We define $P$ to mark the point in time where the light curve power law slope is $\alpha = \alpha_0 + P \times (\alpha_1 - \alpha_0)$, or in other words when a fraction $P$ (e.g. 0.90 or 0.50) of the steepening is obtained. The associated time $\tau_P$ now follows from solving
\begin{equation}
\frac{\dev \log F}{\dev \log \tau} = P \alpha_1 + (1 - P) \alpha_0
\end{equation}
for $\tau$, where $F$ the Beuermann fit function as defined by eq. \ref{Beuermann_fit_function_equation}. A direct measure of the transition duration is provided by $\tau_P / \tau_b$, which has the simple analytical form
\begin{equation}
\tau_P / \tau_b = \left[ (1 - P) / (P) \right]^{1 / (\Delta \alpha \sigma)},
\end{equation}
where $\Delta \alpha \equiv \alpha_1 - \alpha_0$.

Applying this measure to the fit results tabulated in tables \ref{powerlaw_fit_optical_table} and \ref{powerlaw_X-rays_table} we find that the transition duration is very short, typically on the order of a few at most. For $\nu < \nu_c$, $\tau_P / \tau_b$ is essentially independent of synchrotron slope $p$, with differing in the range $p = 2 \ldots 3$ at most by around a single percent. For $\nu > \nu_c$, the differences between different $p$ values are somewhat larger, with on-axis differences up to 20 percent. We have tabulated the average values, weighed in the same manner as $\tau_b$. Given its weak dependence on $p$, $\tau_P / \tau_b$ is arguably a more insightful measure of the nature of the jet break than $\sigma$. It also allows for a direct comparison with earlier estimates by \cite{Kumar2000}. Based on analytical modeling, these authors estimate a transition duration of about a decade in time, contradicted by our simulation-based results (see also the discussion in \citealt{Granot2007}, where it is demonstrated that different analytical transition duration predictions are very sensitive to the precise model assumptions). From an observational perspective, our numerical results are consistent with e.g. the findings of \cite{Zeh2006}, supporting the notion that at least some of the pre-swift bursts discussed by these authors contain jet breaks for explosion in a homogeneous medium.

\subsection{Implications for fit functions}

A comparison of the $\chi^2$ fit results for the different fit functions shows that the performance of the different functions is comparable. Smooth power law fits of type \emph{sB} by definition outperform sharp power law fits, since the latter are a special case of the former, with $\sigma \to \infty$. Smooth power law fits of type \emph{sH} perform poorly for on-axis observers, but often slightly outperform the other fit functions for off-axis observers, which is remarkable since fit function \emph{sB} has more free parameters. Which fit function to use in practice will depend on the goal of the fit. If the goal is to obtain a fit as close to the data points and with as few parameters as possible, \emph{sH} is a good starting point. On the other hand, if the aim is to derive model parameters from the data for the type of model discussed in this paper, \emph{sB} or even sharp power laws (\emph{PL}) might be preferable, since especially for observers close to the axis, their $\alpha_0$ values lie consistently closer to theoretical expectations (and model input for the synthetic light curves).

\section{Summary and discussion}
\label{summary_section}

In this paper we present light curves for gamma-ray burst afterglows decelerating into a constant density circumburst medium. These light curves have been calculated from high-resolution AMR RHD simulations on a grid that is given a Lorentz boost in the direction of the jet, relative to the origin of the explosion. The advantage of this approach is that the relative Lorentz factors in the outflow are reduced and Lorentz contraction of the shock front no longer presents a numerical resolution issue when blast wave deceleration at early times is calculated. The added complexity introduced by the loss of simultaneity across the moving grid relative to the rest frame of the burster can be dealt with by local inverse Lorentz transformations. A linear radiative transfer approach to synchrotron emission through the evolving fluid as represented by a large number of data dumps from the simulation \citep{vanEerten2009BMscalingcoefficients, vanEerten2010transrelativistic} is still possible, as has been presented in 
this paper.

The dynamics of narrow and ultra-relativistic jets will be discussed in \cite{MacFadyen2013}. In the current study we focus on the radiation and the nature of the observed jet break. In a given asymptotic spectral regime, the shape of the light curve is completely determined by the scale-invariant evolution of the spectral breaks and the peak flux. The functions describing these evolutions are characteristic functions of observer angle $\theta_{obs}$ and initial jet half-opening angle $\theta_0$ only and can be scaled between different explosion energies and circumburst densities. Since they are also independent of synchrotron accelerated particle slope $p$, they can be used to generate light curves for arbitrary value of $p$. Generalized scaling relations for arbitrary circumburst density profiles (including ISM and stellar wind) are provided.

The time evolutions of the spectral breaks and peak flux change directly following the jet break, although, thanks to the vast improvement in resolution, an earlier reported temporary post-break steepening of the cooling break $\nu_c$ is found to have been resolution-induced. Nevertheless, the temporal behavior of $\nu_c$ for off-axis observers was found to be extremely sensitive to small deviations from radial flow, even at early times and this is likely to leave an inprint in observations, although any specific model (such as a structured jet, \citealt{Meszaros1998, Rossi2002, Kumar2003, Granot2005}) prediction might be hard to disentangle from the effects of changes in the synchrotron emission process (see e.g. \citealt{Filgas2011}).

The shape of the jet break is systematically surveyed for jet opening angles $\theta_0 = 0.05$, $0.1$, $0.2$ rad., observer angles ranging from on the jet axis to on-edge and $p$ values ranging from 2.01 to 3.0. Pre-break temporal indices are found to be in good agreement with theoretical expectations for purely radial flow. This is partially a consistency check on the computer code, since purely radial BM flow was used to set up the initial conditions of the simulations. On the other hand, the simulations were started from an ultra-relavistic on-axis Lorentz factor $\gamma_0 = 100$ and minor deviations from radial flow will therefore have occurred well before the jet break. 

For the cases considered, post-break temporal indices are generally far steeper than theoretically expected for a quickly expanding jet. This does not imply exponential jet expansion actually occurred \citep{vanEerten2012observationalimplications, MacFadyen2013} as demonstrated by the dependency of the jet break shape on the observer angle, but represents the combined effect of expansion and the edges of the outflow becoming visible. The difference in slopes between the synthetic light curves and those reported for the \emph{Swift} sample \citep{Racusin2009} means that it is exceedingly difficult, at least for the \emph{Swift} sample and at least for on-axis observers, to reconcile the data with the model of an initially top-hat blast wave decelerating into a constant density medium. Even for off-axis observers this is becoming problematic, although a number of caveats apply: the jet break might be simply post-poned beyond what \emph{Swift} can observe \citep{vanEerten2010offaxis, vanEerten2011hiddenswift}, or only a fraction of an off-axis jet break is seen. Sharp power law fits confirm that the jet break time is sensitive to the observer angle and increases significantly as the observer moves off-axis, which has implications for the interpretation of afterglow data and inferred energy of the explosion (which will be overestimated when an on-axis observer is assumed, as discussed in \citealt{vanEerten2010offaxis}).

This discrepancy between light curve slopes from ISM simulations and \emph{Swift} (or other instrument) data can in theory be explained by assuming that afterglow blast waves decelerate instead into a stellar wind environment shaped by the progenitor star. The most likely scenario then is one where a jet in a stellar wind environment is viewed almost on-edge, given a random orientation of the jet. The jet-break is generally less steep for a stellar wind environment \citep{Kumar2000, Granot2007, DeColle2012stratified}. A further complication is added by the fact that GRB progenitor stars are not expected to exist in complete isolation, and the stellar wind environment of the star is likely to be shaped by multiple colliding stellar winds \citep{Mimica2011}. Full results for the stellar wind case computed from a boosted frame will be presented in a follow-up study. Alternatively, the explosion does occur in a homogeneous medium but the jet break is hidden from view by an additional physical process, such as prolonged injection of energy into the blast wave (see e.g. \citealt{Nousek2006, Panaitescu2006, ZhangBing2006, Panaitescu2012}).

Different power law fit functions have been used in the literature to describe jet breaks. Comparing sharp power laws, smoothly connected power laws (`\emph{sB}', \citealt{Beuermann1999}) and power law transitions including an exponential term (`\emph{sH}', \citealt{Harrison1999}), we find that all descriptions provide good fits to synthetic light curves, although type \emph{sH} underperforms for on-axis observers and often outperforms the other types for off-axis observers. Nevertheless, type \emph{sB} fit functions and sharp power laws yield pre-break results that are the easiest to interpret in terms of the underlying model.

The simulation data, light curves and characteristic functions (i.e. the scale-invariant time behavior of peak flux and spectral breaks) in this work will be used to improve the accuracy of simulation-based data fitting methods such as \textsc{boxfit} \citep{vanEerten2012boxfit}. From the characteristic functions light curves for each asymptotic spectral regime can be reproduced directly, as has been done in this paper and following the approach from \cite{vanEerten2012scalings}. For the full spectrum, a heuristic description of the sharpness of spectral transitions is required as well (see \citealt{Granot2002, Leventis2012} for an example of this approach in the spherical case). Alternatively, the simulation output for the blast wave dynamics can be processed using the methods employed for \textsc{boxfit}, albeit with the extra step of transforming to the lab frame. This has the advantage that radiative transfer equations can subsequently be performed very quickly and that no heuristic description of the spectral transitions is needed.

As stated earlier, the steepness of the post-break slopes poses a challenge for the \emph{Swift} sample. A true test of the severity of this issue is to compare observational data and synthetic light curves systematically using one of the simulation-based fit approaches described above. This will be the topic of future work. We note that the one afterglow that has already been fitted using the \textsc{boxfit} approach, GRB 990510, has a steep post-break temporal slope compared to those in the \emph{Swift} sample ($F \propto t^{-2.40}$, according to \citealt{Stanek1999}), which helps to explain how it was possible to obtain a good fit using the ISM model for that particular burst.

All light curves and spectral break and peak flux evolution functions from this work will be made publicly available on-line at \url{http://cosmo.nyu.edu/afterglowlibrary}

\acknowledgments

This research was supported in part by NASA through grant NNX10AF62G issued through the Astrophysics Theory Program, by the NSF through grant AST-1009863 and by Chandra grant TM3-14005X. Resources supporting this work were provided by the NASA High-End Computing (HEC) Program through the NASA Advanced Supercomputing (NAS) Division at Ames Research Center. The software used in this work was in part developed by the DOE-supported ASCI/Alliance Center for Astrophysical Thermonuclear Flashes at the University of Chicago.

\appendix

\section{Light curves from a boosted frame}
\label{boosted_frame_appendix}

As in previous work \citep{vanEerten2010transrelativistic, vanEerten2010offaxis}, the radiative transfer equation is solved in the burster frame (the ``lab'' frame) simultaneously for a large number of rays through the evolving fluid. Because the simulation grid is itself moving with a fixed Lorentz factor, the simulation frame is no longer equal to the lab frame. The consequence of this is that additional Lorentz transformations will be necessary going from simulation to burster frame, not only to boost the fluid quantities, but also to take into account the loss of burster frame simultaneity across a single snapshot. As a consequence of the latter, the contributions from a single snapshot to the emission and absorption coefficients of the rays, for a given observer time and angle, no longer lie on a flat intersecting plane (previously labeled `equidistant surface' or `EDS'), but on a curved surface. Denoting observer time $t_{obs}$, burster time $t$ and simulation grid time $t'$, we have for each ray on 
each snapshot the 
constraint
\begin{equation}
t_{obs} = t(t') - R(t'), 
\end{equation}
where $R$ the distance traveled by the ray in the burster frame parallel to the line of sight. $R$ is equal to zero when the ray crosses the EDS plane centered on the burster frame origin and oriented perpendicular to the line of sight (i.e. defined such that light emitted from the origin at $t = 0$ will arive at observer time $t_{obs} = 0$). This plane, which we label `EDS0', is defined in the burster frame and therefore still flat.). For any given ray, the relevant coordinate for a given snapshot is
\begin{equation}
 \vec{q} = \vec{q}_E + R \hat{u}_E = \vec{q}_E + (t - t_{obs}) \hat{u}_E,
\end{equation}
where $\vec{q}_E$ the coordinates of the point where the ray crosses EDS0 and $\hat{u}_E$ a unit vector pointing along the ray to the observer. Writing the vector components of the previous equation explicitly, we get
\begin{equation}
 \left( \begin{array}{c} q_x \\ q_y \\ q_z \end{array} \right) = \left( \begin{array}{c} q_{Ex} \\ q_{Ey} \\ q_{Ez} \end{array} \right) + (t - t_{obs}) \left( \begin{array}{c} \sin \theta_{obs} \\ 0 \\ \cos \theta_{obs} \end{array} \right).
\label{coordinates_equation}
\end{equation}
Before the radiative transfer calculations are performed, we pre-process the snapshot files to store the local fluid states in terms of burster frame coordinates $(\vec{q}, t)$. The relevant Lorentz boost equations for a boost of factor $\gamma_S$ and velocity $\beta_S$ along $z$, the direction of the jet, are
\begin{eqnarray}
 t & = & \gamma_S (t' + \beta_S q_z' ), \nonumber \\
 q_z' & = & \gamma (q_z - \beta_S t).
\end{eqnarray}
Combining these with equation \ref{coordinates_equation} allows us to determine which fluid cell to probe for a given ray (determined by its $\vec{q}_E$ coordinates) and given snapshot (determined by its simulation frame time $t'$), leading to:
\begin{eqnarray}
 q_z & = & \frac{q_{Ez}}{1 - \beta_S \cos \theta_{obs}} + \frac{c t' \cos \theta_{obs}}{\gamma_S (1 - \beta_S \cos \theta_{obs})} - \frac{c t_{obs} \cos \theta_{obs}}{1 - \beta_S \cos \theta_{obs}}, \nonumber \\
 t & = & \frac{t'}{\gamma_S} + \frac{\beta_S q_z}{c}, \nonumber \\
 q_y & = & q_{Ey} \nonumber \\
 q_x & = & q_{Ex} + c (t - t_{obs} ) \sin \theta_{obs}.
\end{eqnarray}
The distance $\dev R$ traveled by each ray between two snapshots that are $\dev t'$ apart is given by
\begin{equation}
 \dev R = c \dev t = \frac{c \dev t'}{\gamma_S (1 - \beta_S \cos \theta_{obs})}.
\end{equation}
The local emission and absorption coefficients are a function of \emph{comoving} fluid number density $n$, \emph{comoving} fluid energy density $e$ and burster frame fluid velocity $\vec{v}$. The comoving quantities are provided directly by the fluid simulation, since they are independent of the grid velocity. The velocity and Lorentz factor in the burster frame are calculated during the pre-processing of the grid snapshots according to the standard relativistic velocity addition rules.

\bibliography{jetbreakshape}

\end{document}